\documentclass[twocolumn]{article}

\usepackage[utf8]{inputenc} 
\usepackage[T1]{fontenc}

\usepackage{amsmath}
\usepackage{amssymb}
\usepackage{amsfonts}
\usepackage{mathrsfs}
\usepackage{mathtools}
\usepackage{relsize}

\usepackage{graphicx}
\usepackage{dcolumn}
\usepackage{bm}
\usepackage{xcolor}
\usepackage{tikz}
\usepackage[normalem]{ulem}
\usepackage{hyperref}

\hyperref[<label>]{}

\usepackage[left=0.75in,right=0.75in,top=1in,bottom=1in,%
            footskip=.25in]{geometry}

\usepackage{background}
\backgroundsetup{
  position=current page.east,
  angle=-90,
  nodeanchor=east,
  vshift=-4mm,
  hshift=38mm,
  opacity=1,
  scale=3,
  contents=\textsc{PREPRINT}\ \ -- \ \ \textsc{PREPRINT} \ \ -- \ \ \textsc{PREPRINT}
}

\definecolor{mygreen}{rgb}{0., 0.66667, 0.2666} 
\definecolor{mygreendark}{rgb}{0., 0.502, 0.} 
\definecolor{mypurple}{rgb}{0.83137, 0., 0.58824} 
\definecolor{myorange}{rgb}{1., 0.4, 0.} 

\newcommand{\greenline}{\raisebox{2pt}{\tikz{\draw[-,black!0!mygreen,solid,line width = 1.2pt](0,0) -- (5.5mm,0);}}}
\newcommand{\greenlinedark}{\raisebox{2pt}{\tikz{\draw[-,black!0!mygreendark,solid,line width = 1.2pt](0,0) -- (5.5mm,0);}}}
\newcommand{\redline}{\raisebox{2pt}{\tikz{\draw[-,black!0!red,solid,line width = 1.2pt](0,0) -- (5.5mm,0);}}}

\newcommand{\bluedashedline}{\raisebox{2pt}{\tikz{\draw[-,black!0!blue,dashed,line width = 1.2pt](0,0) -- (5.3mm,0);}}}
\newcommand{\blackdashedline}{\raisebox{2pt}{\tikz{\draw[-,black!100!blue,dashed,line width = 1.2pt](0,0) -- (5.3mm,0);}}}

\newcommand{\orangedashedline}{\raisebox{2pt}{\tikz{\draw[-,black!0!myorange,dashed,line width = 1.2pt](0,0) -- (5.3mm,0);}}}

\newcommand{\blacklinelongdash}{\raisebox{2pt}{\tikz{\draw[-,black!100!blue,line width = 1.2pt](0,0) -- (2.2mm,0); \draw[-,black!100!blue,line width = 1.2pt](2.8mm,0) -- (5mm,0);}}}

\newcommand{\purpledottedline}{\raisebox{2pt}{\tikz{\draw[-,black!0!mypurple,dotted,line width = 1.2pt](0,0) -- (5.2mm,0);}}}

\newcommand{\bluedashdottedline}{\raisebox{2pt}{\tikz{\draw[-,black!0!blue,dash dot,line width = 1.2pt](0,0) -- (5.0mm,0);}}}

\newcommand{\blackcircle}{\raisebox{1.0pt}{\tikz{\draw[black,line width = 1.0pt](0,0) circle (1.5pt);}}}
\newcommand{\bluetriangle}{\raisebox{1.0pt}{\tikz{\draw[blue,line width = 1.0pt](0,0) -- (0.1cm,0) -- (0.05cm,0.1cm) -- (0,0);}}}

\begin{document}

\twocolumn[
  \begin{@twocolumnfalse}

    \textbf{PREPRINT - Submitted to Physics of Plasmas.}

    \begin{center}
      {\Large \textbf{Collisionless ion modeling in Hall thrusters:\\[1ex]
analytical axial velocity distribution function and heat flux closures}}
    \end{center}

    {\normalsize Stefano Boccelli$^{1,\,a)}$, Thomas Charoi$^{2,\,b)}$, Alejandro Alvarez-Laguna$^{2,\,c)}$, Pascal Chabert$^{2,\,d)}$, Anne Bourdon$^{2,\,e)}$ and Thierry E.\ Magin$^{3,\,f)}$}

    \vskip2ex

    {\footnotesize
    $^{1}$ Department of Aerospace Science and Technology, Politecnico di Milano, Milano, Italy.\\
    $^{2}$ Laboratoire de Physique des Plasmas, \'Ecole Polytechnique, Palaiseau, France.\\
    $^{3}$ Aeronautics and Aerospace Department, von Karman Institute for Fluid Dynamics, Sint-Genesius-Rode, Belgium.}

    \vspace{2ex}

    {\footnotesize
    $^{a)}$ stefano.boccelli@polimi.it\\
    $^{b)}$ thomas.charoy@lpp.polytechnique.fr\\
    $^{c)}$ alejandro.alvarez-laguna@lpp.polytechnique.fr\\
    $^{d)}$ pascal.chabert@lpp.polytechnique.fr\\
    $^{e)}$ anne.bourdon@lpp.polytechnique.fr\\
    $^{f)}$ magin@vki.ac.be}
    

    \vskip2ex
    \textbf{PREPRINT - Submitted to Physics of Plasmas.}

    \begin{abstract}
    The genesis of the ion axial velocity distribution function (VDF) is analyzed for collisionless Hall thruster discharges.
    An analytical form for the VDF is obtained from the Vlasov equation, by applying the Tonks-Langmuir theory in the thruster channel, under the simplifying assumptions of monoenergetic creation of ions and steady state.
    The equivalent set of 1D unsteady anisotropic moment equations is derived from the Vlasov equation, and simple phenomenological closures are formulated, assuming a polynomial shape for the ions VDF.
    
    The analytical results and the anisotropic moment equations are compared to collisionless PIC simulations, employing either a zero heat flux (Euler-like equations) or the polynomial-VDF closure for the heat flux.
    The analytical ion VDF and its moments are then compared to experimental measurements.
    \vspace{1cm}
    \end{abstract}

  \end{@twocolumnfalse}
]

\section{Introduction}

\begin{figure}[htb]
  \centering
  \includegraphics[width=0.9\columnwidth]{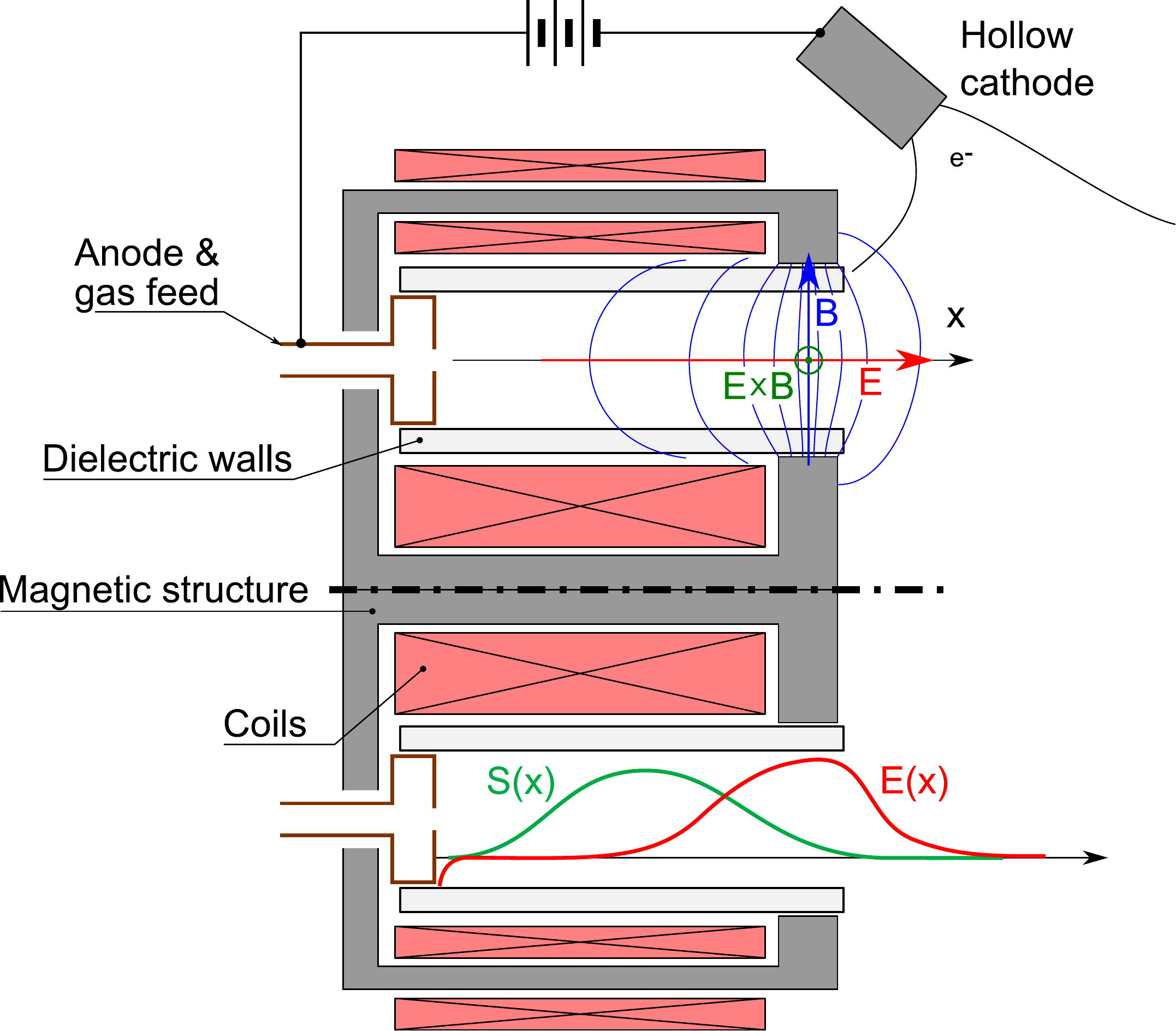}
  \caption{Schematic view of a Hall thruster, with axial electric field $E(x)$ and ionization profile $S(x)$. The radial magnetic field $B$ is mostly concentrated at the channel exit plane.}
  \label{fig:hall-thruster-schematic}
\end{figure}

Hall thruster devices, also known as stationary plasma thrusters,\cite{zhurin_physics_1999} are a class of electric space propulsion devices, with very high efficiency and specific impulse, compared to chemical thrusters.\cite{goebel2008fundamentals}
A stream of neutral gas is injected from a perforated anode (see Fig.~\ref{fig:hall-thruster-schematic}) and is ionized through collisions with free electrons emitted by an externally mounted hollow cathode.
The presence of an external radial magnetic field strongly limits the mobility of electrons from the cathode to the anode, creating a region of high axial electric field, near the exit plane of the thruster.
This crossed configuration for the electric and magnetic fields forces electrons to drift along the azimuthal direction.
The bulk of ion production, located inside the channel, is generated by the impact of hot electrons (at a temperature of some eV, and comparable kinetic energy in the azimuthal direction) with the cold and slow neutrals.
Ions, which are substantially unmagnetized due to their large mass, are then accelerated by the axial electric field up to velocities of the order of $15-20$ km/s.\cite{ahedo_plasmas_2011}

From their earliest developments in the `60s, Hall thrusters have now reached a mature stage, and are currently used aboard many satellites.\cite{goebel2008fundamentals}
However, their numerical modeling has proved to be challenging due to the need of considering a number of phenomena, including plasma waves and instabilities, plasma-surface interaction and chemical and kinetic non-equilibrium.\cite{boeuf_tutorial:_2017}
The lack of numerical simulation tools able to be at the same time accurate and efficient, proves to be an issue for what concerns the development process. This problem especially concerns the scaling of these devices, which is tackled by long and costly experiments or using simplified correlations,\cite{dannenmayer_elementary_2011} but also regards lifetime predictions due to wall erosion.\cite{manzella_predicting_2004}

\subsubsection*{Kinetic non-equilibrium}

The low number of gas-phase collisions and the presence of a strong electromagnetic field enhance kinetic non-equilibrium in Hall thrusters.\cite{taccogna_non-equilibrium_2016}
The resulting non-Maxwellian velocity distribution functions (VDFs) determine the transport properties of the plasma, and are likely to influence the appearance and saturation mechanisms of plasma instabilities\cite{lafleurPSST2018} observed both experimentally\cite{Tsikata2009,Cavalier2013} and numerically.\cite{Adam2004}

\begin{figure}[htb]
  \centering
  \includegraphics[width=1.0\columnwidth]{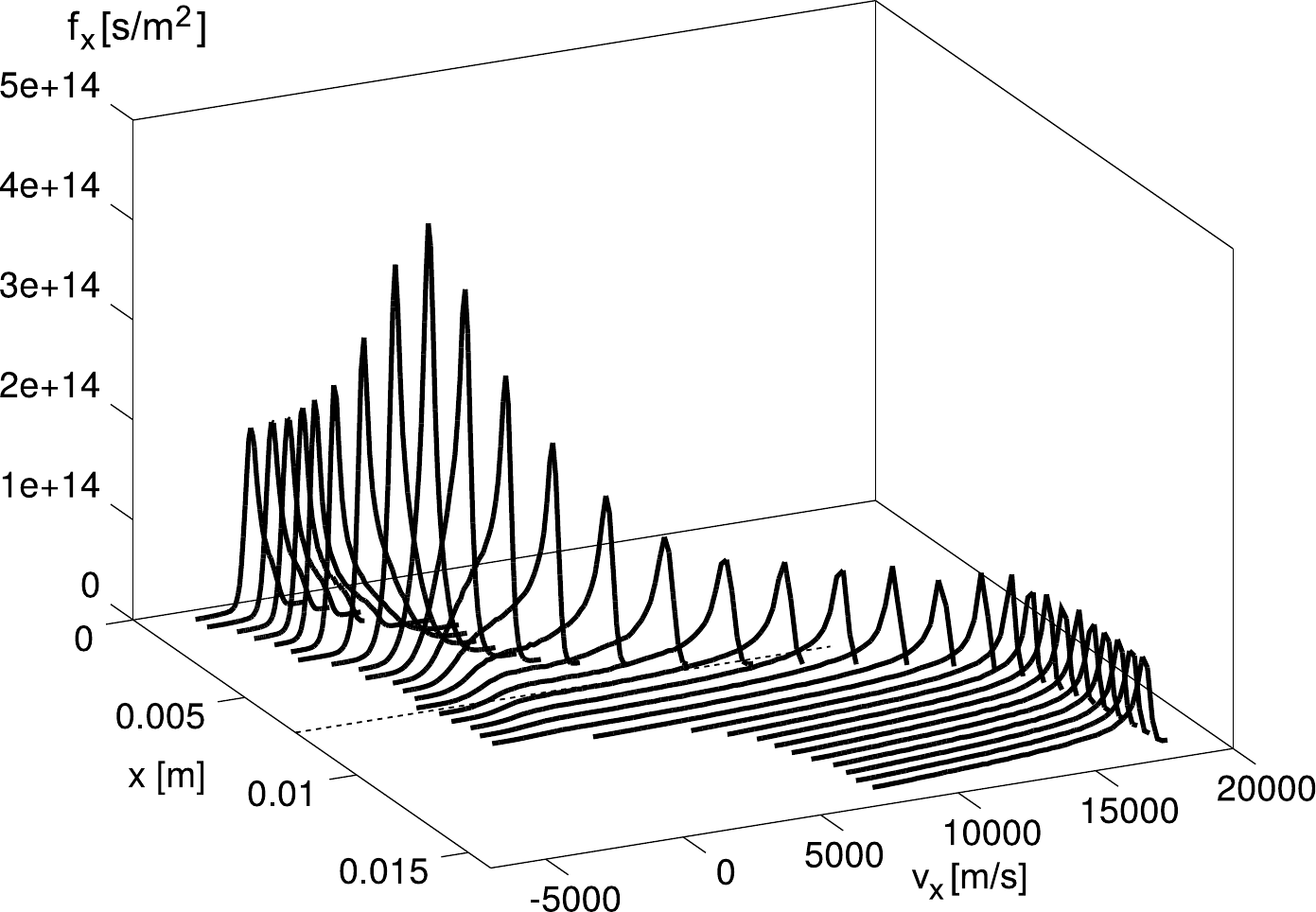}
  \caption{Ions axial distribution function $f_x(x, v_x)$ from a PIC simulation of a Hall thruster discharge (conditions from test case A, in Appendix \ref{sec:appendix-A}). 
  The anode is located at $x = 0\ \mathrm{m}$, and the cathode is out of the shown domain, at $x = 0.025\ \mathrm{m}$.
  Position $x = 0.0075$ m marks the thruster exit plane.}
  \label{fig:VDF-PIC-3d}
\end{figure}

For ions in particular, the importance of elastic collisions can be shown to be small if compared to the accelerating electric field, as can be seen by analysis of the non-dimensional numbers characterizing the problem.
Once an ion is generated inside the channel by an electron-neutral ionizing collision, it is accelerated towards the exit, with little further interaction with neutrals and electrons.
The ions velocity distribution function has been discussed thoroughly in the literature, both from the experimental and numerical perspectives,\cite{hara2012one,Mazouffre2013} and its highly non-Maxwellian shape confirms the secondary role of collisions for this species.
It has often been observed\cite{garrigues_computed_2012,king_ion-energy_2004} that the ion axial VDFs are composed by a dominant peak, followed by a plateau at the lower velocities, or a slowly decaying tail, as shown in the Particle-In-Cell (PIC) computation of Fig.~\ref{fig:VDF-PIC-3d}.

Low collisionality results in a lack of thermalization mechanisms. 
While the axial momentum and energy increase under the effect of the electric acceleration, no relaxation can occur with the radial and azimuthal components. 
This generates a strong anisotropy in the temperature and pressure fields.
Other interactions and phase-space mixing mechanism can arise due to plasma waves developing in the thruster,\cite{boeuf_e_2018,Croes2017} but they will be neglected in the current work.
Transport quantities, such as the heat flux, are also heavily affected by non-equilibrium.

Given the degree of kinetic non-equilibrium, a proper framework for describing Hall thruster discharges is the Vlasov equation,\cite{van_kampen_theoretical_1967} as opposed to reduced order moment descriptions obtained by integrating microscopic properties in the velocity space.
However, the high dimensionality of the Vlasov formulation (3 spatial plus 3 velocity dimensions), together with the time-step constraints imposed by plasma oscillations makes kinetic simulations very expensive.
Currently, two-dimensional PIC simulations take a few days to a few months on cluster architectures, and the only option for moving to 3D simulation still consists in using scalings of geometrical or physical constants,\cite{taccogna_three_2011} but does not guarantee a complete similitude of the problem.
This computational complexity plays in favor of simplified fluid descriptions, based for example on the solution of lower order moments of Vlasov's equation.\cite{barral_numerical_2001}
Such systems suffer however from the lack of a proper closure for the transport quantities, such as the heat flux, and the degree of non-equilibrium that can be obtained (and thus the accuracy of the method) is only as good as the closure itself.

The commonly employed cold-ions approximation\cite{ahedo_one-dimensional_2001} consists in neglecting the ions pressure and solving only the mass and momentum conservation equations.
This approach allows to capture reasonably well the ion density and velocity fields.
However, it does not provide any information on the ion temperature, which is suggested to be linked to the appearance of azimuthal instabilities.\cite{Lafleur2017}
The situation is improved by introducing an energy equation, and the resulting system is often closed by assuming a zero heat flux,\cite{barral_numerical_2001} leading to Euler-like equations.
This improves the prediction of the velocity field by adding a pressure gradient term in the momentum equation, but still the prediction of higher order moments is not necessarily accurate.
%
Classical fluid dynamic approaches towards a moderate non-equilibrium description, such as the Navier-Stokes-Fourier (NSF) equations, lose their validity in the Hall thruster regimes, as far as ions are concerned.
Indeed, the low collisionality and the presence of a strong electric acceleration make perturbative solution methods such as the Chapman-Enskog method (on which the NSF equations are based) theoretically invalid, as the distribution function cannot be assumed to be a small perturbation of a local Maxwellian.\cite{ferziger_mathematical_1972}

\subsubsection*{Aim and structure of this work}

This work only considers collisionless ions moving in the axial direction.
\begin{itemize}
    \item first, we analyze in detail the axial ion distribution function, and provide a simplified analytical formulation, valid in steady and quasi-steady state conditions;
    \item then, an unsteady non-equilibirum fluid formulation is derived for ions and its solution is compared to more comprehensive kinetic formulations.
\end{itemize}

The first point is addressed as follows.
Section \ref{sec:IVDF} aims at analyzing how the ion velocity distribution function arises from the interplay between the electric field and the ionization profile along the channel.
A simple analytical solution is derived in Section \ref{sec:analytical-VDF}, following the treatment for the ion VDF in collisionless plasma sheaths.\cite{tonks_general_1929,harrison_low_1959,ingram_ion_1988}
A number of assumptions are introduced at this stage, to obtain a simple analytical expression.
Moments of the analytical VDF are obtained in Section \ref{sec:moments-analyt-vdf}.

Then, in Section \ref{sec:anisotropic-fluid-formulation} attention is moved towards developing a fluid-like model for the said problem.
In Section \ref{sec:anisotropic-fluid-eqs}, an anisotropic fluid model is formulated starting from the Vlasov equation.
A number of assumptions introduced in Section \ref{sec:analytical-VDF} are relaxed, and the resulting system of equations can describe the axial dynamics of ions in both the steady and unsteady regimes.
A closure for the resulting system of equations is discussed in Section \ref{sec:polynomial-phen-closure}, where a phenomenological form of the heat flux is introduced, which mimics the observed features of the analyzed distribution functions by use of low-order polynomials.

Finally, the analytical solution and the system of fluid equations are tested on four test cases, in Section \ref{sec:results}.
In the first three test cases, we compare our results to 2D collisionless PIC simulations, for conditions typical of Hall thrusters (discussed in Appendix \ref{sec:appendix-A}). This allows for a detailed comparison of the distribution function and its first four moments.
In the last test case, we compare our results to experimental measurements, with the aim of providing a partial validation of our analytical model.

As mentioned, the present work assumes that ions are collisionless.
This limits its validity to the region of strong electric field inside the thruster, and at most in the very first part of the plume expansion.
The effect of collisions is briefly discussed in Section \ref{sec:results}.
All throughout the work, we only consider singly charged ion species (although a generalization is trivial), treating the electric field and ionization profiles as imposed quantities.
The models developed can be directly applied to fully self-consistent multi-fluid simulations, where the electric field and ionization rate are obtained from the simulation at each integration step.


\section{Genesis of ions VDF}\label{sec:IVDF}

To describe the shape of the axial ion distribution function, we consider steady state conditions. 
We assume that ions are produced only by electron-neutral collisions, during which the heavy species velocity can be approximated to be unchanged.
Ions are thus created at the local velocity distribution of neutrals, which in this section is assumed to be a monoenergetic beam at velocity $v_n$. 
Additionally, $v_n$ will be considered uniform along the channel.
The effect of this assumption may be argued to quickly become small, if compared to the electrostatic acceleration along the channel.
The same stands for the effect of a more reasonable model for ions injection (such as a drifted Maxwellian centered on the neutrals velocity $v_n$, in place of the current monoenergetic model, see Appendix \ref{sec:appendix-B}).
These assumptions will be relaxed in the fluid formulation of Section \ref{sec:anisotropic-fluid-eqs}.

\begin{figure}[htb]
  \centering
  \includegraphics[width=1.0\columnwidth]{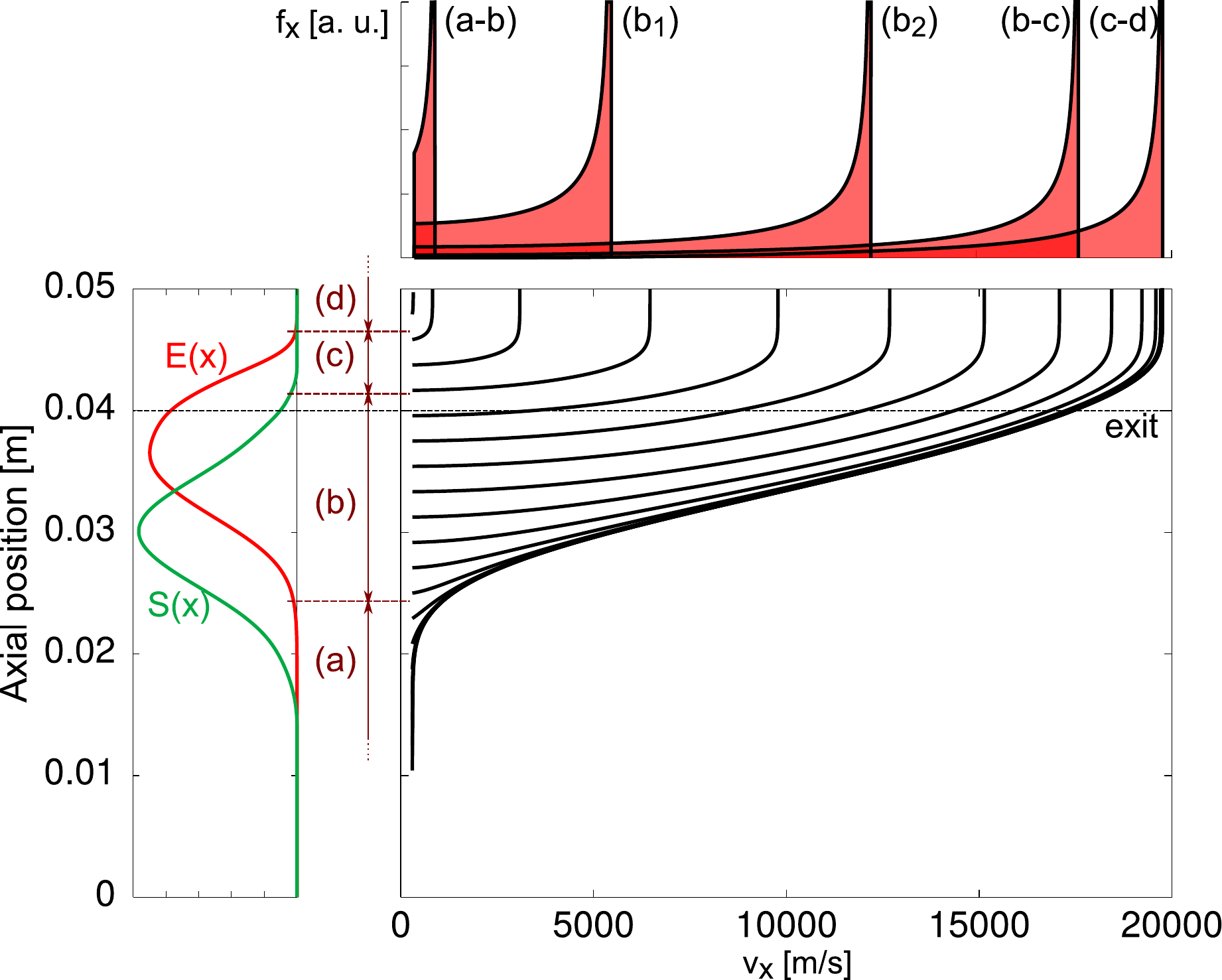}
  \caption{Phase-space trajectory of ions (bottom-right), obtained by direct integration of the motion inside the electric field, and identification of sub-regions, together with ionization source term $S(x)$ and electric field $E(x)$ profiles (left), and resulting distribution functions at selected locations (top), arbitrarily scaled. Insert: zoom on the ion trajectories inside region (a).
  Thruster exit plane is at $x=0.04 \ \mathrm{m}$.}
  \label{fig:phase-space-traj-E-S}
\end{figure}

Ion creation and acceleration mechanisms are conveniently analyzed in the phase space.
For the sake of the present discussion, let us consider as an example the case of Fig.~\ref{fig:phase-space-traj-E-S}, where we show typical values for Hall thrusters.
In this example, the thruster exit plane is located at $x=0.04$ m and we assume an injection velocity in the order of $v_n = 300 \, \mathrm{m/s}$.
The electric field and ionization axial profiles of this example were adapted from Boeuf and Garrigues, \cite{boeuf_low_1998} with maximum electric field $E$ around 20 kV/m and maximum ionization rate $S$ of $2.5\times 10^{23}$ ions/s/m$^3$.

We start by considering the simple case of a positive electric field all throughout the domain (no region of velocity inversion for ions).
Ions originate at position $v_n$ along the velocity axis, and their generation along the thruster channel typically starts before the location of the maximum electric field.
Referring to the case of Fig.~\ref{fig:phase-space-traj-E-S},
the first ions are created after position $x \approx 0.01$ m.
We mark this as the beginning of Region (a).
In this first part, the electric field is often quite low.
As a result, the ions concentration builds up in a tiny volume of phase space and trajectories are almost superimposed.
Ions slowly drift towards the exit mainly due to their (low) initial velocity $\approx v_n$, and gradually accelerate.
Indeed, the phase-space trajectory is initially almost vertical and very little acceleration occurs until the electric field begins to be significant.
The ions produced in this region will have the highest velocity at the exit of the thruster, as they can exploit the full length of the acceleration region, and will constitute the high velocity peak in the VDF.

As the electric field increases in {Region (b)}, trajectories become steeper.
The ionization source term is significant in this region, or even peaking.
This region exhibits a continuous strong production of ions, whose velocity at the exit will be lower or at most equal to that of ions originating in {Region (a)}, due to the shorter acceleration region available.
The ions generated in this region will form the body of the final VDF (as observed for example by Mazouffre and Bourgeois\cite{mazouffre_spatio-temporal_2010}).
At this location the VDF looks like the distribution $\mathrm{b_1}$, in Fig.~\ref{fig:phase-space-traj-E-S}-top.
The body, initially resembling a plateau, gradually transforms into a long-decaying tail as the ionization profile decreases ($\mathrm{b_2}$).

Moving towards the thruster exit plane, the ionization profile becomes negligible.
We denote this as the beginning of {Region (c)}.
Ions are still accelerated by a considerable electric field, but there is no more production at the velocity $v = v_n$, which results into a mere translation (with some deformation as well) of the distribution function towards higher velocities.
This region may include the last part of the acceleration region and/or the first part of the plume.

Finally, we denote the bulk of the thruster plume as {Region (d)}.
As both the electric field and the ionization source have become negligible, the effect of collisions gradually becomes the leading term in the ion dynamics.
We will not study the effect of collisions, thus limiting the validity of the model to Regions (a), (b) and (c).


\subsection{Analytical ion VDF}\label{sec:analytical-VDF}

Under the simplifying hypotheses aforementioned, it is possible to obtain an analytical solution for the ion axial distribution function.
The derivation follows the classical results for plasma sheaths.\cite{tonks_general_1929,harrison_low_1959,ingram_ion_1988}
We start from the Vlasov equation for ions, neglecting the magnetic field, with electric field $\bm{E}(\bm{x})$ and ionization source term in phase-space $\mathcal{S}(\bm{x},\bm{v})$, with $\bm{x}$ and $\bm{v}$ the space and velocity coordinates:
\begin{equation}\label{eq:vlasov-equation}
  \frac{\partial f}{\partial t} + \bm{v} \cdot \frac{\partial f}{\partial \bm{x}} + \frac{q \bm{E}}{m} \cdot \frac{\partial f}{\partial \bm{v}} = \mathcal{S}(\bm{x},\bm{v})
\end{equation}

\noindent where $q$ and $m$ are the ion charge and mass respectively.
In this section, we consider the steady state case only, imposing $\partial_t f \equiv 0$.
In the most general case of a multi-species description of a gas/plasma, the ionization source term $\mathcal{S}$ is an integral operator, accounting for the reaction cross-sections and the distribution functions of reactants and products.\cite{giovangigli_multicomponent_2012}
Since in the present case we are considering ions only, the term $\mathcal{S}$ simplifies considerably.
Assuming ions are created as a monoenergetic beam with velocity $\bm{v} = (v_n, 0, 0)$, the ionization source term becomes: $\mathcal{S}(\bm{x}, \bm{v}) = S(x)\,\delta(v_x - v_n) \,\delta(v_y) \,\delta(v_z)$, where $S(x)$ is expressed in $[\mathrm{s}^{-1}\mathrm{m}^{-3}]$ and is imposed along the axis in this work.
Additionally, in the case of a purely axial electric field, and neglecting its variations along the azimuthal and radial directions $y$ and $z$, we have: $\bm{E} = E(x) \, \hat{x}$. 
Under such assumptions, the solution of Eq.~\ref{eq:vlasov-equation} is purely one-dimensional, and we can drop all dependence on the $y$ and $z$ variables. 

The presence of azimuthal instabilities for the electric field would break the present assumptions.
However, such limitation is less severe than it may seem, as it still allows to retrieve reliable results, in terms of azimuthal averages.
This will be confirmed by results in Section~\ref{sec:results}.
The assumption of radial symmetry is also questionable in real Hall thrusters due to the presence of walls. 
Such assumption limits the validity of our model to the region of the channel center-line.

In absence of collisions and relaxation mechanisms, and under the said symmetry assumptions, the three components of particle motion are decoupled.
In order to restrict our attention to the axial motion of particles ($v_x$), we integrate the Vlasov equation over the $v_y$ and $v_z$ velocity components, obtaining an equation for the marginal distribution function $f_x$, in one space and one velocity dimensions (1D1V):
\begin{equation}
  v_x \frac{\partial f_x}{\partial x} + \frac{q E}{m} \frac{\partial f_x}{\partial v_x} = S(x) \, \delta(v_x - v_n)
\end{equation}

\noindent where $f_x$ is defined as:
\begin{equation}\label{eq:marginal-VDF-fx}
  f_x(v_x) = \int \!\! \int_{-\infty}^{+\infty} \!\! f(\bm{x},\bm{v}) \ \mathrm{d} v_y \, \mathrm{d} v_z
\end{equation}

\noindent where we omitted the dependence of $f_x$ in the $x$ spatial coordinate for lighter notation. 
From a given profile of the electric field and production term $S(x)$, the solution is easily found
by following characteristic lines, corresponding to particle trajectories in the 1D1V phase space.
In the collisionless case, ions generated at the position $x_0$ fall freely along the electric potential $\phi(x)$, such that their velocity at position $x$ will be:
\begin{equation}\label{eq:velocity-potential}
  v_x(x_0, x) = \left[ \tfrac{2q}{m} \left( \phi(x_0) - \phi(x) \vphantom{\sum}\right) + v_n^2 \right]^{\frac{1}{2}}
\end{equation}

\noindent For each location $x$ along the thruster, this equation maps the ions with axial velocity $v_x$ to the location $x_0$ where they were generated.
As it appears from Fig.~\ref{fig:phase-space-traj-E-S}, the particles generated between $x_0$ and $x_0+\mathrm{d}x$ will have at position $x$ a velocity between
$v_x(x_0, x)$ and $v_x(x_0, x)+\mathrm{d}v$, and the following balance of fluxes holds:\cite{harrison_low_1959} 
\begin{equation}\label{eq:fluxes-balance}
  v_x f_x(v_x)\, \mathrm{d} v_x = -S(x_0) \, \mathrm{d} x_0
\end{equation}

\noindent where the minus sign accounts for the inverse relation between a growing $x_0$
and its corresponding final velocity.
As the velocity $v_x$ of a free-falling ion is known in function of $x_0$, one computes:
\begin{equation}
    \tfrac{\mathrm{d} v_x}{\mathrm{d} x_0} 
  = 
    \tfrac{\mathrm{d} }{\mathrm{d} x_0} 
      \left[ \tfrac{2q}{m} \left( \phi(x_0) - \phi(x) \right) + v_n^2 \right]^{\frac{1}{2}}  
  = -\tfrac{q}{m} \tfrac{E(x_0)}{v_x(x_0; x)}
\end{equation}

\noindent and inserting it into the Eq.~\eqref{eq:fluxes-balance}, one gets the simple analytical expression for the distribution function:
\begin{equation}\label{eq:analytic-VDF}
    f_x(v_x(x_0)) \ = \ - \frac{S(x_0)}{v_x} \frac{\mathrm{d} x_0}{\mathrm{d} v_x}
     \ = \
    \frac{m}{q} \frac{S(x_0)}{E(x_0)}
\end{equation}

\noindent In the steady case, it is thus possible to know exactly the ions VDF if $S$ and $E$ are known, for example at a given step of a numerical simulation, or from measurements inside the thruster.

\begin{figure}[htpb]
  \centering
  \includegraphics[width=\columnwidth]{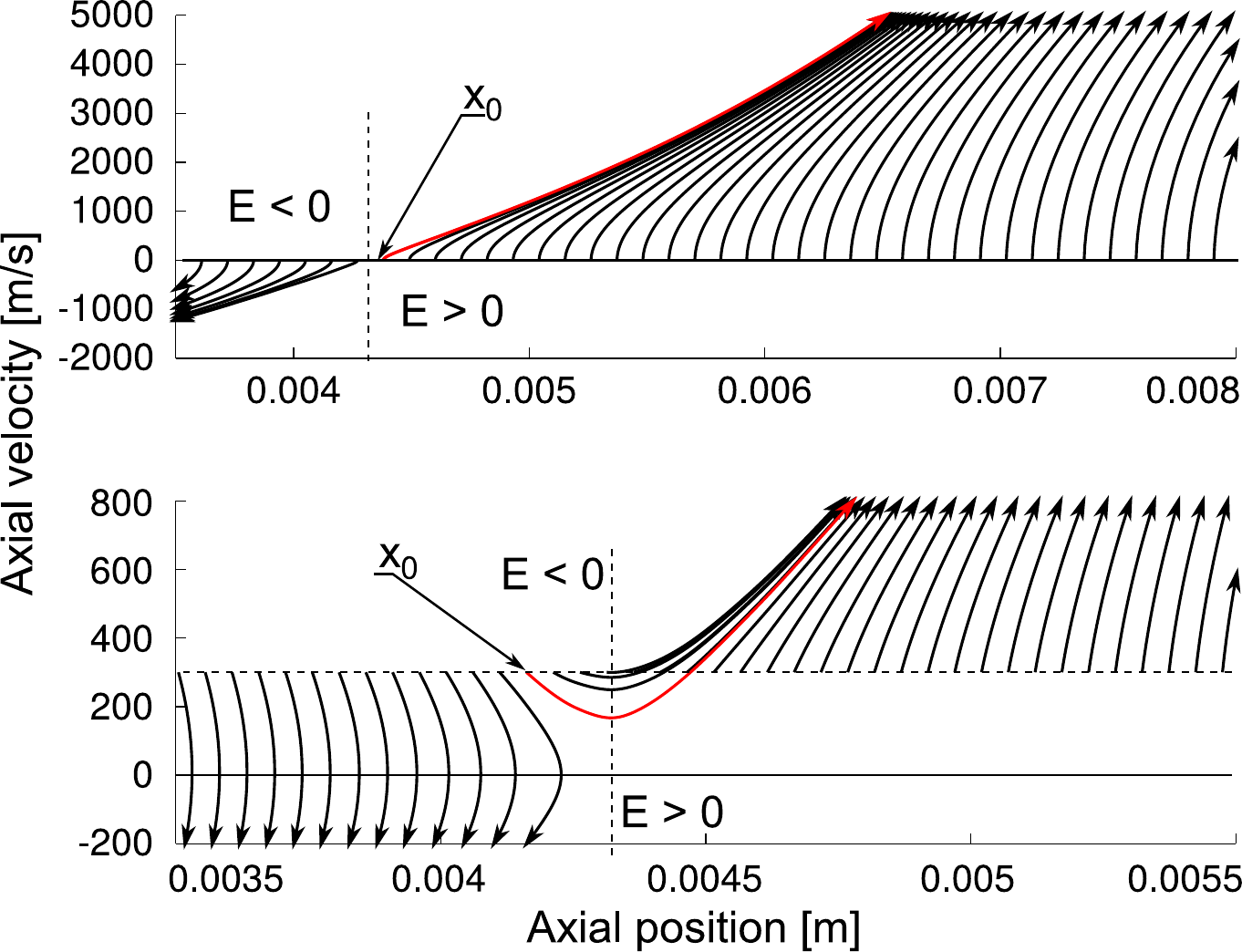}
  \caption{Typical phase space trajectories around a point of electric field inversion. Top: zero injection velocity $v_n = 0$; Bottom: $v_n = 300$ m/s.
  Parameter $x_0$ is identified by the first particle trajectory reaching the considered location (red line for $x>0.0045$ m).}
  \label{fig:example-x0-traj}
\end{figure}

Practically speaking, first of all the desired location $x$ is chosen, at which the VDF is to be plotted.
Then a vector of values for $x_0$ is created (where $x_0 < x$ for positive electric fields), which are used to sample the values of $S(x_0)$ and $E(x_0)$.
Finally, the values of $v_x$ which correspond to locations $x_0$ are obtained from Eq.~\eqref{eq:velocity-potential}.

\begin{figure*}[htpb]
  \centering
  \includegraphics[width=\textwidth]{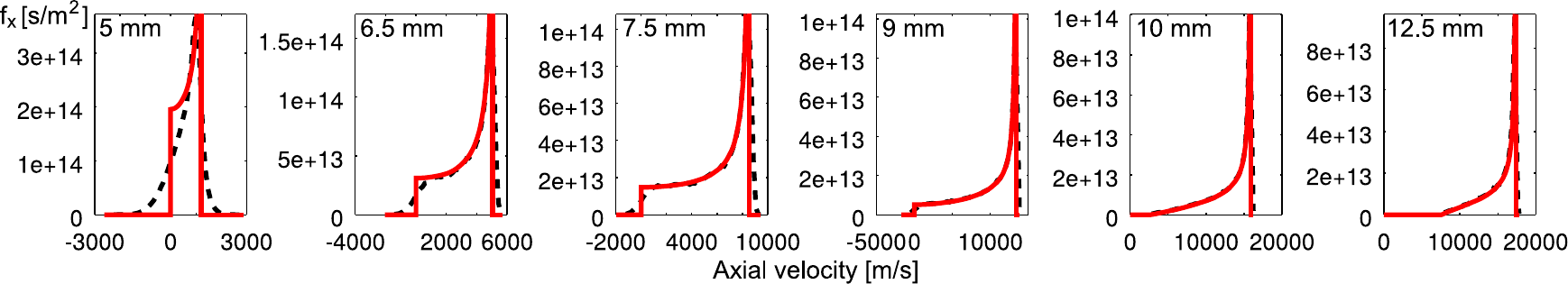}
  \caption{Analytical ({\protect\redline}) and  numerical ({\protect\blackdashedline}, particle-in-cell) velocity distribution function, at various locations along the Hall thruster discharge. The conditions are these of test case A.}
  \label{fig:testcase-A-VDF-compar}
\end{figure*}

If the electric field is positive all along the channel, $x_0$ simply coincides with the beginning of the domain.
In case the electric field changes direction along the channel, the choice of $x_0$ requires some more discussion.
A point of electric field inversion indeed exists inside Hall thruster channels, and is located near the anode.
Considering the phase-space trajectories of Fig.~\ref{fig:example-x0-traj}, obtained from a changing-sign electric field, one can see that for a given position $x$, the particles that reach this location and thus contribute to the distribution function are only those generated between $x_0$ and $x$.
Particles generated for $\tilde{x} < x_0$ backflow towards the left and do not contribute to the VDF at position $x$.
The most intuitive possibility consists in inferring $x_0$ graphically, from the trajectories plot.
In case of a zero injection velocity, $x_0$ simply coincides with the nodal point of the electric field (Fig.~\ref{fig:example-x0-traj}-top).
For a general injection velocity $v_n > 0$, some additional ions created in the region of slightly negative electric field can escape the electric field barrier (Fig.~\ref{fig:example-x0-traj}-bottom).
The value for $x_0$ is to be chosen upstream of the nodal point for the electric field, considering that the ion birth kinetic energy can overcome an additional potential difference.
In this way, all the ions contributing to the VDF are accounted for.

From a practical perspective, the prediction of the acceleration region could be in most cases performed effectively by neglecting the effect of $v_n$ and directly taking $x_0$ at the point where $E = 0$.

As an example, Fig.~\ref{fig:testcase-A-VDF-compar} compares the analytical velocity distribution function to the numerical Particle-In-Cell VDF that was already shown in Fig.~\ref{fig:VDF-PIC-3d} (test case A, also see Appendix \ref{sec:appendix-A}).
As expected, we observe a distribution that gradually evolves into a beam-like distribution as ions are accelerated along the channel and reach the plume, with a long low-energy tail.
The error present at locations of low kinetic energy (around $x=5$ mm) comes from assuming a monoenergetic ions birth. 
However, as soon as the acceleration starts, the importance of injection details quickly vanishes, and the numerical VDF is reproduced with fidelity.

\subsection{Moments of the analytical VDF}\label{sec:moments-analyt-vdf}

Thermodynamic quantities can be readily obtained from a distribution function, by computation of its moments in velocity space.\cite{ferziger_mathematical_1972}
We shall recall here the kinetic definition of some low order moments.
The number density $n$ for instance is a zero-th order moment, coming from the direct integration of the distribution function.
Considering then quantities specific of the axial direction $x$, that is the focus of this work, the average velocity $u_x$ in the $x$ direction is obtained from the momentum $n u_x$, obtained from the first order moment of the VDF.
We denote by $P_x$ the first component of the pressure tensor $P_{ij}$, describing the flux of momentum due to the thermal motion of particles.
This is closely linked to $e_x$, the thermal energy per unit mass associated with the axial motion of particles.
Finally, we define the heat flux $Q_x$ as (half of) the first component of the heat flux tensor $Q_{ijk}$.
\begin{subequations}
  \begin{align}
    n(x)   &= \mathsmaller{\int_{-\infty}^{+\infty}} f(v) \, \mathrm{d}^3 v \label{eq:n-kinetic-def}\\
    n(x)\, u_x(x)   &= \mathsmaller{\int_{-\infty}^{+\infty}} v_x f(v) \, \mathrm{d}^3 v \label{eq:ux-kinetic-def}\\
    P_x(x) &= \mathsmaller{\int_{-\infty}^{+\infty}} m (v_x - u_x)^2 f(v) \, \mathrm{d}^3 v \label{eq:P-kinetic-def} \\
    n(x)\, e_x(x) &= \mathsmaller{\int_{-\infty}^{+\infty}} \tfrac{1}{2} (v_x - u_x)^2 f(v) \, \mathrm{d}^3 v \label{eq:ex-kinetic-def} \\
    Q_x(x) &= \mathsmaller{\int_{-\infty}^{+\infty}} \tfrac{m}{2} (v_x - u_x)^3 f(v) \, \mathrm{d}^3 v \label{eq:Qx-kinetic-def}
  \end{align}
\end{subequations}

These moments can be evaluated for the analytical solution obtained in Section \ref{sec:analytical-VDF} (superscript ``$a$''), considering the definition of the marginal distribution $f_x(v_x)$, in Eq.~\eqref{eq:marginal-VDF-fx}, and employing the fluxes balance in Eq.~\eqref{eq:fluxes-balance}.
The analytical moments are here given as integral expressions along the $x$ axis, and can be readily evaluated numerically from the values of $S(x)$ and the potential $\phi(x)$.
The number density reads:
\begin{equation}\label{eq:analytical-n}
  n^a(x)
= 
  \int_{x_0}^x \frac{S(x_0)}{v(x_0; x)} \,  \mathrm{d} x_0
\end{equation}

\noindent where function $v(x_0,x)$ is defined in Eq.~\eqref{eq:velocity-potential}.
The average axial velocity $u_x^a(x)$ is obtained from:
\begin{equation}\label{eq:analytical-u}
  u_x^a(x) 
= 
  \frac{1}{n^a}\int_{x_0}^x S(x_0) \, \mathrm{d} x_0
\end{equation}

\noindent The pressure term $P^a_x(x)$ can be computed once
the average velocity $u_x^a$ is known:
\begin{equation} \label{eq:analytical-Px}
  P^a_x(x) = \int_{x_0}^x \!\!m \, \frac{S(x_0)}{v(x_0; x)} \left[v(x_0; x) - u^a_x(x)\right]^2 \, \mathrm{d} x_0
\end{equation}

\noindent and the heat flux term $Q^a_x(x)$ reads:
\begin{equation} \label{eq:analytical-Qx}
  Q^a_x(x) = \int_{x_0}^x \frac{m}{2}\frac{S(x_0)}{v(x_0; x)} \left[ v(x_0; x) - u^a_x(x)\right]^3\, \mathrm{d} x_0
\end{equation}

It should be recalled here that these ``analytical'' moments suffer from the very same assumptions of the analytical distribution function, namely (i) ions are injected as a monoenergetic beam, neglecting the birth temperature, (ii) are collisionless, and (iii) in steady state.
The choice for the lower integration extreme $x_0$ was detailed in Section \ref{sec:analytical-VDF}.


\section{Anisotropic fluid formulation}\label{sec:anisotropic-fluid-formulation}

For multi-dimensional cases, a full kinetic description of Hall thruster geometries requires very high computational efforts.
On the other hand, reduced order descriptions such as fluid models can be much lighter and thus allow for agile evaluations of the thruster performance, and for iterative design procedures.
However, the accuracy of classical fluid formulations (such as the Euler or the Navier-Stokes-Fourier equations) strongly depends on the closure employed, and their validity for highly non-equilibrium cases is thus questionable.

A sound fluid description for ions in Hall thrusters should account for the non-equilibrium in the distribution function, in particular in terms of anisotropy of energies, resulting from the low number of collisions and strong acceleration in the axial direction.
A proper treatment for the heat flux also needs to be developed, providing a reasonable closure for the system of equations.

In this section, we first formulate governing equations for the anisotropic case, describing the mass, the axial momentum and the energy associated to the axial motion of ions.
Then, we close these equations by formulating a phenomenological heat flux closure based on a prescribed polynomial form for the VDF.


\subsection{Anisotropic fluid equations}\label{sec:anisotropic-fluid-eqs}

Fluid-like equations for ions in the axial direction $x$ are obtained as moments of the Vlasov equation, Eq.~\eqref{eq:vlasov-equation} (see for example Ferziger and Kaper\cite{ferziger_mathematical_1972} and Benilov\cite{benilov_kinetic_1997}).
We avoid the commonly employed ``cold ions approximation'',\cite{ahedo_one-dimensional_2001} as it may include significant errors in the momentum.
The amount of error can be estimated by comparing the contributions of $\rho u_x^2$ and $P_x$
obtained from the analytical solution (reaching errors of 15-20\% for the test cases of Section \ref{sec:results}).
Instead we solve an equation for the ion energy.

As ions are weakly collisional, we choose a fully anisotropic description.
Equations are written for the axial component of the momentum and for the energy associated to the axial velocity of particles alone.
The quantities for the other two directions evolve separately and can thus develop different values for the pressure and temperature.
In absence of collisions or electromagnetic coupling terms, no relaxation term appear.

Integrating the Vlasov equation weighted by the microscopic property $\psi$, we obtain the generalized 
moment equation:\cite{ferziger_mathematical_1972}
\begin{equation}\label{eq:general-eq-moments}
  \frac{\partial \, n\, \overline{\psi}}{\partial t}
+ \frac{\partial}{\partial \bm{x}} \cdot \left[ \ n \, \overline{\bm{v} \psi}  \ \right]
=
  \frac{n q\bm{E}}{m} \cdot \overline{\frac{\partial \psi }{\partial \bm{v}}}
+ \left.\overline{\frac{\delta \psi}{\delta t}} \right|_{\mathrm{c}} 
\end{equation}

\noindent with the definitions for the operator $\overline{\bullet}$ and for the chemical production source:
\begin{subequations}
  \begin{align}
    \overline{\bullet \vphantom{phi}} &\equiv \frac{1}{n} \int_{-\infty}^{\infty} \bullet \, f \ \mathrm{d}^3 v \\
    \left.\overline{\frac{\delta \psi}{\delta t}}\right|_{\mathrm{c}} \!\!
 &\equiv \frac{1}{n} \int_{-\infty}^{\infty} \psi S(x,v) \, \mathrm{d}^3 v
  \end{align}
\end{subequations}

In the present derivations, we perform the same spatial symmetry assumptions of Section~\ref{sec:analytical-VDF}: space derivatives along $y$ and $z$ are dropped. 
The former condition leads to a formulation to be interpreted as an average along the azimuthal direction, and the latter limits the validity of the model to the region of the channel centerline.
The mass, x-momentum and axial energy equations are obtained choosing respectively $\psi$ equal to the mass of ions, the x-component of the momentum $m v_x$ and the axial energy $m v_x^2/2$.
The first component of the pressure tensor $P_{x}$ is used in place of the average axial energy $m n e_x$. The two are linked by their kinetic definition: $P_x \equiv 2\, m n e_x$, as can be seen from Eqs.~\eqref{eq:P-kinetic-def} and \eqref{eq:ex-kinetic-def}, such that the total axial energy becomes $\rho E_x = (\rho u_x^2 + P_x)/2$.
Notice that this form for the energy coincides with the classical gas dynamic definition ``$\rho u^2/2 + P/(\gamma - 1)$'', where the adiabatic constant $\gamma$ is taken equal to 3, describing a monatomic gas with a single translational degree of freedom, as in the current case.
The equations take the form:
\begin{equation}\label{eq:conservation-eq-general}
  \frac{\partial \bm{U}}{\partial t} + \frac{\partial \bm{F}}{\partial x} = \bm{G}
\end{equation}

\noindent where the vector of variables in conservative form $\bm{U}$ and their fluxes vector $\bm{F}$ read:
\begin{equation}\label{eq:U-F-anisotr}
  \bm{U} =
  \begin{pmatrix}
    \rho \\
    \rho u_x \\
    \tfrac{1}{2}\left(\rho u_x^2 + P_x\right)
  \end{pmatrix}
\ , \ \
  \bm{F} =
  \begin{pmatrix}
    \rho u_x \\
    \rho u_x^2 + P_x \\
    \tfrac{1}{2}\rho u_x^3 + \tfrac{3}{2} u_x P_x + Q_x
  \end{pmatrix}
\end{equation}

\noindent and with the RHS source terms:
\begin{equation}\label{eq:S-anisotr}
  \bm{G} =
  \begin{pmatrix}
    m S \\
    n q E + R_x^r \\
    n q E u_x + C_x^r
  \end{pmatrix}
\end{equation}

\noindent where $\rho = m n$ is the mass density, and $S = S(x)$ the imposed ionization term.
It is worthwhile to stress that while these equations describe only the axial motion of particles, they could be easily coupled to equations for the radial and azimuthal components.

The relation between temperature and pressure can be easily written by considering the axial thermal energy per unit mass $e_x = k_B T_x / 2 m$ and from definition in Eq.~\eqref{eq:P-kinetic-def} we retrieve: $P_x = n k_B T_x$, where $T_x$ is the axial temperature.

Terms $R_x^r$ and $C_x^r$ represent sources for the momentum and energy equations
due to the production of ions at a given initial momentum and energy respectively.
For ions produced from a Maxwellian population of neutrals at temperature $T_n$ and centered around an axial velocity $v_n$, these terms read:
\begin{equation}
  R_x^r = S \,m v_n \ \ \ \ \ \ \mathrm{and} \ \ \ \ \ \ C_x^r = S \, \left( \frac{m_n v_n^2}{2}  + \frac{k_B T_n}{2}\right)
\end{equation}

\noindent The term $k_B T_n/2$ arises from considering motion along one axis only (and would be $3/2 \,k_B T$ in a fully isotropic formulation, including the azimuthal and radial particles energy, in equilibrium at temperature $T$).
The terms $v_n$ and $T_n$ are to be imposed in the present equations, and can be a function of the position along the channel.
Note that this gives additional flexibility with respect to the previous analytical solution, where we had assumed $v_n$ uniform in space and $T_n = 0$ (Section \ref{sec:analytical-VDF}).

The equations formulated require a closure, obtained by expressing the heat flux term $Q_x$ 
in function of the available moments $\rho$, $u_x$ and $P_x$.
The simplest closure consists in choosing, arbitrarily enough, that $Q_x=0$.
This results in the Euler equations, which are theoretically valid 
in the infinitely collisional regime, where distribution functions are Maxwellian, but lose their theoretical validity for  
collisionless and accelerated ions.

This adiabatic closure proves to be reasonably accurate as far as the description of the first two moments is concerned, but as anticipated its accuracy is low in terms of reproducing second- and higher-order moments.
On the other hand, the Fourier closure, commonly employed in fluid dynamics, lacks of physical justification in a fully collisionless context, where additionally the flux of energy is purely governed by the combination of 
electric field and ionization profile.

A simple attempt to overcome these limitations and develop a closure aiming at reproducing the basic kinetic features observed, is provided in the next section.


\subsection{Closures through polynomial VDFs}\label{sec:polynomial-phen-closure}

\begin{figure}[htpb]
  \centering
  \includegraphics[width=0.6\columnwidth]{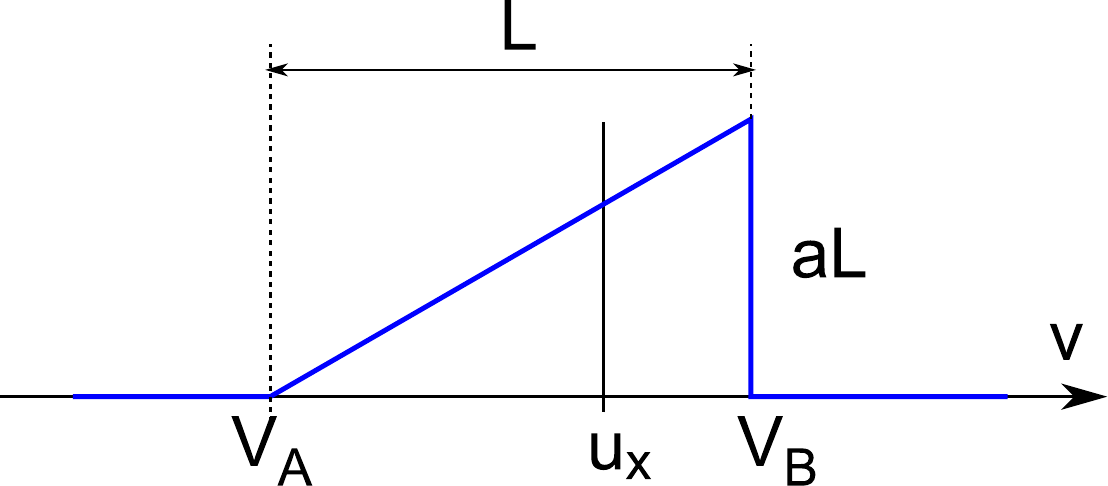}
  \caption{Triangular distribution function ($p=1$) for the heat flux closure.}
  \label{fig:triangular-VDF}
\end{figure}

\begin{figure*}[tb]
  \centering
  \includegraphics[width=\textwidth]{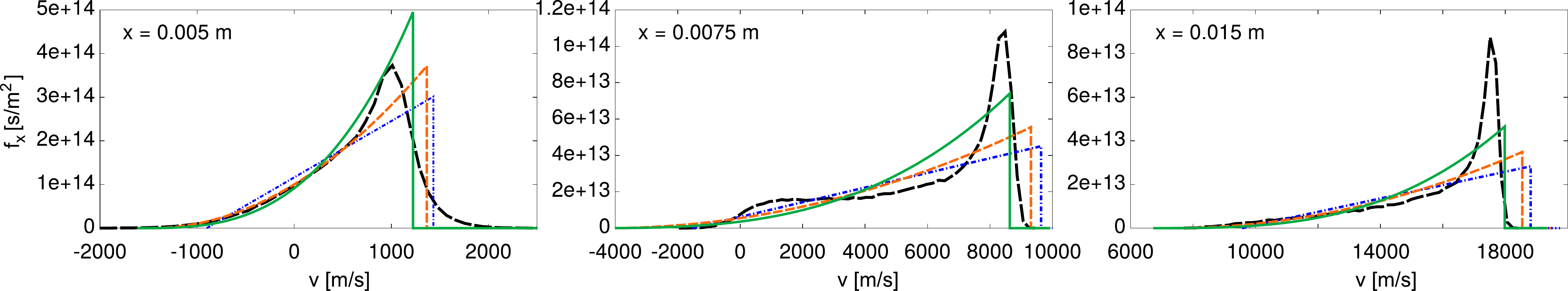}
  \caption{Comparison of PIC and polynomial VDFs with same density, average velocity and temperature, for three selected locations of test case A. PIC VDF ({\protect\blacklinelongdash}), triangular  ({\protect\bluedashdottedline}), parabolic ({\protect\orangedashedline}) and cubic function ({\protect\greenline}) approximations.}
  \label{fig:PIC-vs-polynomial-VDF-shape}
\end{figure*}

In this section we derive a phenomenological closure for the heat flux, inspired by the observed ions VDF. Roughly speaking, the collisionless ion axial VDF consists in a peak accelerated by the electric field, followed by one long plateau or tail. As a third order moment, the heat flux $Q_x$ is driven by the asymmetry of such distribution. As an attempt to mimic this behavior and therefore to reproduce a reasonable heat flux, we assume that the distribution function in the acceleration channel can be represented by a simple polynomial of order $n$, in the form:
\begin{equation}
    f_x (v_x) \approx f^{(p)}(v_x) = 
    \begin{cases}
      a (v_x - V_A)^p &\mathrm{for} \ \ \ v_x \in [V_A, V_B] \\
      0 \ \ \ &\mathrm{otherwise}
    \end{cases}
\end{equation}

\noindent where we omitted the dependence of $f_x$ on the space location for simplicity.
Considering the cases $p=\{1,2,3\}$, we approximate the distribution function by a triangle, a parabola or a cubic function, with support $[V_A, V_B]$.
This distribution is shown in Fig.~\ref{fig:triangular-VDF} for the case of $p=1$, with the definition of the auxiliary parameter $L = |V_B - V_A|$, width of the distribution. 
The other distributions can be seen in Fig.~\ref{fig:PIC-vs-polynomial-VDF-shape}.

Note that through the use of the marginal distribution $f_x(v_x)$, no assumption is being made on the shape of the distribution along the $v_y$ and $v_z$ axes, which can be chosen arbitrarily and does not influence the derivation of purely axial quantities.

The three free parameters of these polynomial distributions, $a$, $V_A$ and $V_B$, can be linked to the density, velocity and temperature of such distributions.
This ultimately allows to write the heat flux $Q_x^{(n)}$ of this distribution as a function of the lower moments, thus providing a closure.
Among the three, the cubic closure showed to provide the best results, therefore calculations will be here provided for the case $p=3$.
Derivations for the cases $p = 1,2$ are analogous and results are reported in Table~\ref{tab:closures-results}.
The closure values for a generic order $p$ is reported in Appendix \ref{sec:appendix-C}.

First, the number density reads:
\begin{equation}\label{eq:n-eq-derivation-cubic}
    n = \int_{-\infty}^{+\infty} f \ \mathrm{d}^3 v
      = \int_{V_A}^{V_B} a (v_x - V_A)^3 \ \mathrm{d} v_x
      = \frac{a L^4}{4}
\end{equation}

Similarly, the average velocity is found from its kinetic
definition:
\begin{multline}
    n u_x = \int_{-\infty}^{+\infty} v f \ \mathrm{d}^3 v = \int_{V_A}^{V_B} \!\!\! v_x\,  a (v_x - V_A)^3 \ \mathrm{d}v_x 
\end{multline}

The integration is easily performed by the change of variables $\xi = v -  V_A$, carrying the integration from $0$ to $L$.
By exploiting Eq.~\eqref{eq:n-eq-derivation-cubic} and the definition of $L$, we find a relation for $V_A$ and $V_B$:
\begin{equation}
    \begin{cases}
      V_A = u_x - 4L/5 \\
      V_B = u_x + L/5
    \end{cases}
\end{equation}

Then, an expression for the distribution width $L$ can be obtained from the temperature definition:
\begin{multline}
   \rho e_x = \frac{P_x}{2} = \frac{n k_B T_x}{2} 
   \equiv \int_{-\infty}^{\infty} \frac{m}{2} (v_x - u_x)^2 f \ \mathrm{d}^3 v \\
   = \int_{V_A}^{V_B} \frac{m a}{2} \, (v_x - u_x)^2 \,(v_x - V_A)^3 \ \mathrm{d}v_x
\end{multline}

\noindent which, with the same change of variables, results in the relation:
\begin{equation}
    L = \sqrt{\frac{75 k_B T_x}{2 m}}
\end{equation}

This completely defines the shape of the polynomial distribution, given the first three moments.
A comparison of some PIC distributions and the polynomial VDFs is shown in Fig.~\ref{fig:PIC-vs-polynomial-VDF-shape} for three selected locations of test case A (see Appendix \ref{sec:appendix-A}).
For the three locations shown, the density, velocity and temperature are obtained from the PIC simulation, and used to compute the polynomial VDF parameters.
The matching shows to be rather approximated, however the heat flux will result to be well reproduced. 
Indeed, reproducing exactly the VDF is often unnecessary in view of obtaining a reasonable value for its lower moments, as many details of the VDF are lost in the integration process for computing the moments.

The heat flux is finally obtained with the same procedure, from the kinetic definition:
\begin{equation}
  Q_x = \int_{V_A}^{V_B} \frac{m a}{2} \, (v_x - u_x)^3 \, (v_x - V_A)^3 \ \mathrm{d} v_x
\end{equation}

\noindent resulting in a closed form for the heat flux:
\begin{equation}
    Q_x = - \frac{2 \, m n}{875} \left(\frac{75 \, k_B T_x}{2\, m} \right)^{3/2}
\end{equation}

Note that since the heat flux is a central moment, the average velocity $u_x$ does not appear in its formulation, but only has the effect of centering the distribution function.

\begin{table}[h]
  \centering
  \begin{tabular}{c|c|c|c}
       &   Triangle  &  Parabola   &   Cubic function \\
     \hline
    $L$ & $\left(18 k_B T_x / m\right)^{1/2}$ &  $\left( 80 k_B T_x / 3 m\right)^{1/2}$ & $\left( 75 k_B T_x / 2 m\right)^{1/2}$ \\
    $a$   & $2 n /L^2$ & $3 n / L^3$ & $4 n / L^4$ \\
    $V_A$ & $u - 2/3 L$ & $u - 3/4 L$ & $u - 4/5 L$ \\
    $V_B$ & $u + 1/3 L$ & $u + 1/4 L$ & $u + 1/5 L$ \\
    \hline
    $Q_x$ & $- m n L^3 / 270$ & $- m n L^3/320$ & $- 2 \, m n L^3/875$\\
  \end{tabular}
  \caption{VDF parameters for polynomial closures for triangle ($p=1$), parabola ($p=2$) and cubic function ($p=3$).}
  \label{tab:closures-results}
\end{table}

\begin{figure}[htb]
  \centering
  \includegraphics[width=0.9\columnwidth]{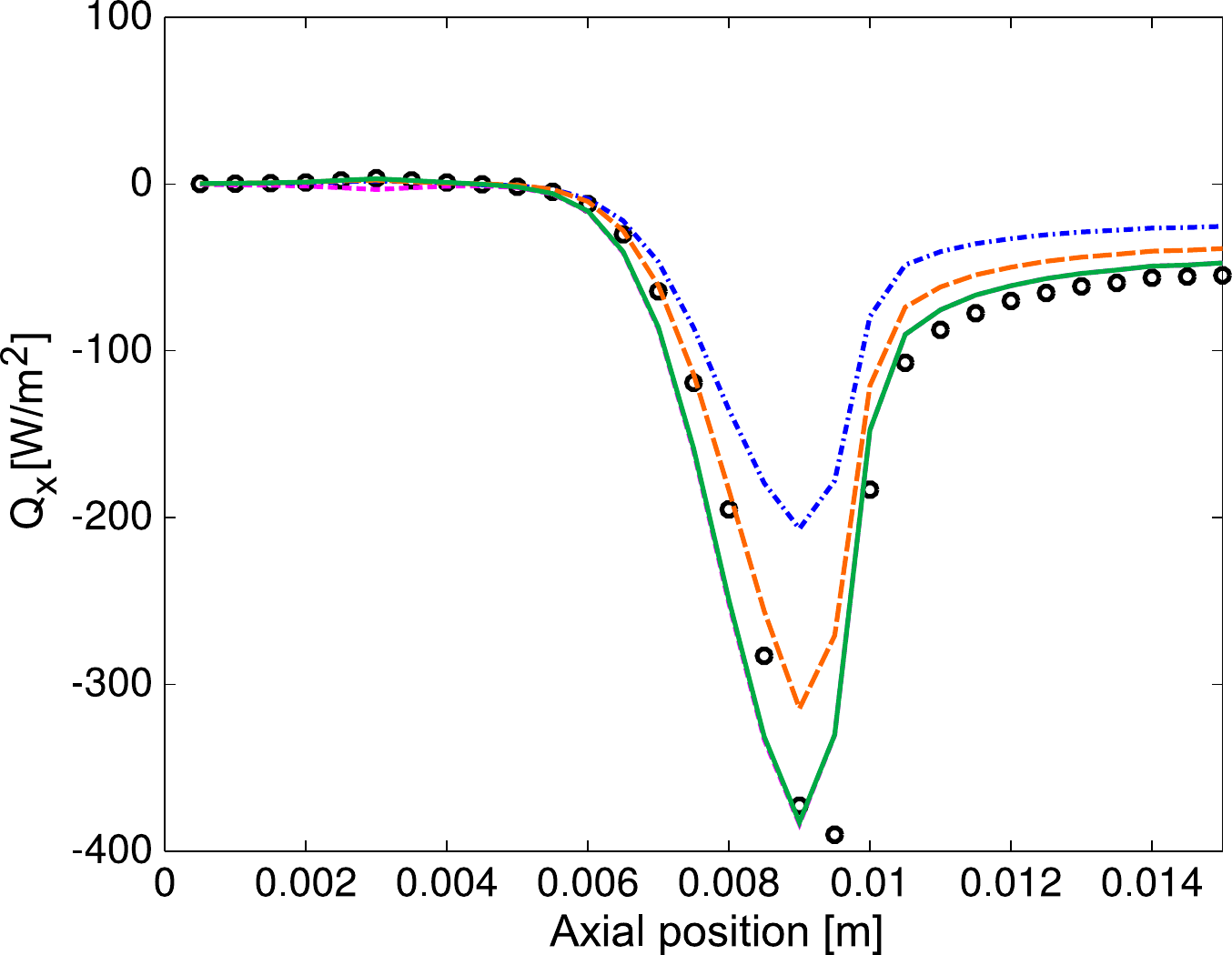}
  \caption{Application of triangular heat flux to the lower moments of the PIC simulation of test case A.
  PIC simulation ({\protect\blackcircle \ \protect\blackcircle \ \protect\blackcircle});
  non-limited triangular ({\protect\bluedashdottedline}), parabolic ({\protect\orangedashedline}) and cubic ({\protect\purpledottedline}) VDF heat fluxes $Q_x$. Cubic closure with $\mathrm{erf}()$ limiting ($Q_x^*$, {\protect\greenline}).   }
  \label{fig:heat-flux-comparison}
\end{figure}

Without needing to solve the full set of fluid equations, we can obtain a preliminary assessment for the validity of the closures by considering the density, velocity and temperature fields from the PIC simulations:
using these fields to compute the parameters in Table~\ref{tab:closures-results}, we can compare the obtained approximated heat fluxes to the self-consistent heat flux from PIC simulation.
This is done in Fig.~\ref{fig:heat-flux-comparison} for the PIC simulation of test case A (Appendix \ref{sec:appendix-A}). 
The actual accuracy of the closure strongly depends on the test case, but shows in all cases at least a good qualitative agreement, with the parabolic and cubic approximations dominating over the triangular one.
Indeed, the triangular distribution is a crude approximation of the actual VDF, and misses both the shape of the low-velocity tail and the location of the high velocity peak (see Fig.~\ref{fig:PIC-vs-polynomial-VDF-shape}-Center and -Right). The parabolic and cubic functions are slightly better in this regard.
A good matching is shown in the acceleration region, especially if compared to the $Q_x = 0$ Euler closure.
An additional correction to the heat flux will be introduced in the next section.

\subsubsection*{Correction for negative and low velocities}

As mentioned earlier, real thruster geometries are characterized by a point of velocity inversion for ions, close to the anode. 
In this region, the electric field becomes negative and attracts the ions, such that the polynomial distributions of Fig.~\ref{fig:triangular-VDF} should be reversed, heading towards negative velocities.
This reflects into a change in the heat flux closure, which should be modified by introducing a $\mathrm{sign}(u_x)$ function.
This however introduces a new issue, namely the heat flux would discontinuously jump from a positive to a negative value across the position of $u_x = 0$.

Moreover, when the average velocity is close to zero (say, roughly lower than the thermal speed), the assumption of a polynomial distribution function becomes very questionable.
In that region, the distribution is indeed closer to the ions birth Maxwellian, since the electric field did not accelerate and deform the VDF yet, and the heat flux is thus zero.
For these two reasons, we introduce an arbitrary limiting on the heat flux in such regions.
By considering the distance between the high velocity extreme $V_B$ and the average velocity $u_x$, we define the quantity $\Delta = |V_B - u_x|$ (resulting in $\Delta = L/3$, $L/4$ or $L/5$ for the triangular,
parabolic and cubic VDFs respectively, where $V_B$ is assumed to be the highest velocity extreme of the distribution, either positive or negative).
We decide to limit the heat flux in the region where $u_x < 2 \Delta$, meaning that the limiting shall apply whenever the average velocity is lower than a percentage of the thermal velocity.
The most obvious choice consists in a linear limiting, defining the linearly corrected heat flux $Q_x^{\mathrm{lin}}$:
\begin{equation}
  Q_x^{\mathrm{lin}} = \begin{cases}
          \mathrm{sign}(u_x) \, \frac{|u_x|}{2\Delta} Q_x \ &\mathrm{if} \ \ |u_x| < 2 \Delta \\
          \mathrm{sign}(u_x) \, Q_x \ \  &\mathrm{otherwise}
        \end{cases}
\end{equation}

\begin{figure}[htb]
  \centering
  \includegraphics[width=0.95\columnwidth]{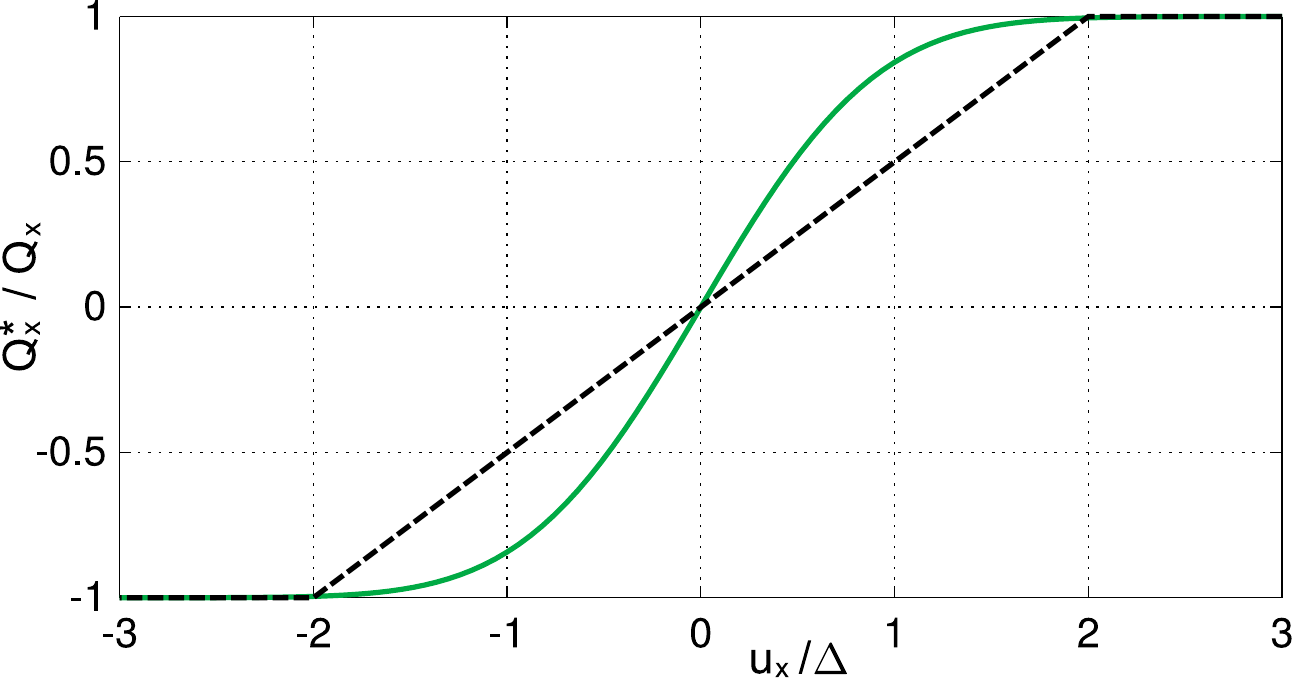}
  \caption{Linear ({\protect\blackdashedline}) and $\mathrm{erf}()$({\protect\greenline}) corrections to the polynomial heat flux.}
  \label{fig:erf-plot}
\end{figure}

However, a non-smooth limiting could introduce some additional numerical difficulties and non-physical behavior in the numerical prediction of the pressure and temperature fields especially.
Therefore, we suggest the use of a smooth sigmoid function, such as the error function $\mathrm{erf}(\chi)$.
Using $\chi = u_x/\Delta$, the $\mathrm{erf()}$ limiting returns the value of $Q_x$ for $u_x \gtrsim 2 \Delta$ (see Fig.~\ref{fig:erf-plot}):
\begin{equation}
    Q_x^* = \mathrm{erf}\left(u_x/\Delta\right) \, Q_x
\end{equation}

\begin{figure}[htb]
  \centering
  \includegraphics[width=0.9\columnwidth]{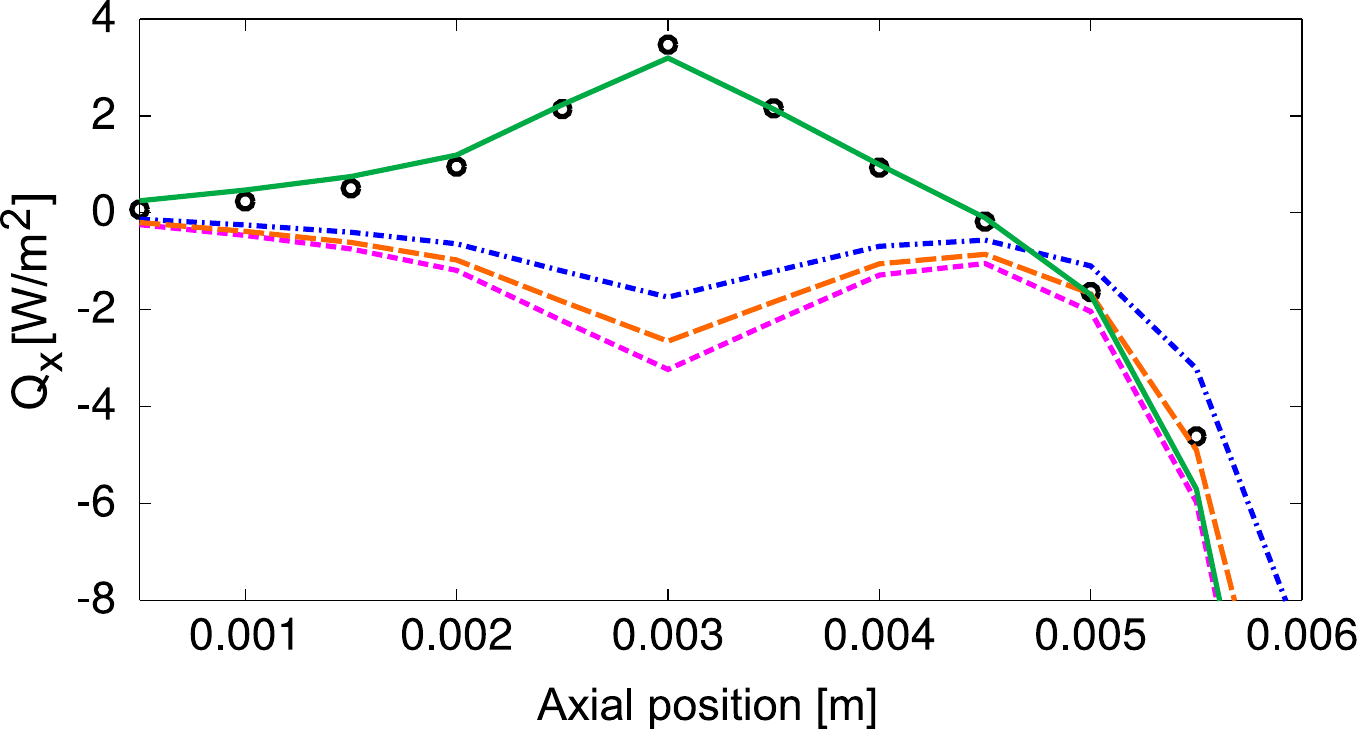}
  \caption{Magnification of Fig.~\ref{fig:heat-flux-comparison} on the posititive heat flux region.
  PIC simulation ({\protect\blackcircle \ \protect\blackcircle \ \protect\blackcircle});
  Non-limited triangular ({\protect\bluedashdottedline}), parabolic ({\protect\orangedashedline}) and cubic ({\protect\purpledottedline}) VDF heat fluxes $Q_x$. Cubic closure with $\mathrm{erf}()$ limiting ($Q_x^*$, {\protect\greenline}).  }
  \label{fig:heat-flux-comparison-zoom}
\end{figure}

This form will be referred to as ``corrected heat flux'' and is the form that we recommend for usage.
Note that it is not necessary to explicitly correct by the sign of $u_x$, being automatically included in $\mathrm{erf}()$.
The simple cubic and this corrected cubic VDF heat flux are shown in Fig.~\ref{fig:heat-flux-comparison}. The corrected closure superimposes on the non-corrected version in the acceleration region, but provides an improved agreement in region where $u_x \approx 0$.
A magnification of Fig.~\ref{fig:heat-flux-comparison} around the region of positive heat flux is provided in Fig.~\ref{fig:heat-flux-comparison-zoom}, where the effect of the correction can be appreciated.

\section{Results}\label{sec:results}

We compare the previous developments against four testcases.

The first three cases are comparisons of the analytical distribution function (Section \ref{sec:results-analytical-vs-PIC}) and the numerical solution of anisotropic fluid equations (Section \ref{sec:results-anisotropic-vs-PIC}) against collisionless PIC simulations.
The PIC simulations are 2D, performed in the axial-azimuthal plane, and describe reasonably well the main features of Hall thruster flows. Details are given in Appendix \ref{sec:appendix-A}. 
Anisotropic equations are closed with the phenomenological cubic-VDF approximation, with $\mathrm{erf()}$ correction.

The fourth test case (Section \ref{sec:experiment-testcase}) shows a comparison of the analytical results to experimental measurements.\cite{mazouffre_spatio-temporal_2010}
We analyze the distribution function, the average velocity and the velocity dispersion, and use this as a tentative verification and comparison for the analytical solution.


\subsubsection{Analytical solution vs PIC - Test cases A, B, C}\label{sec:results-analytical-vs-PIC}

\begin{figure*}[htpb]
  \centering
  \includegraphics[width=1.0\textwidth]{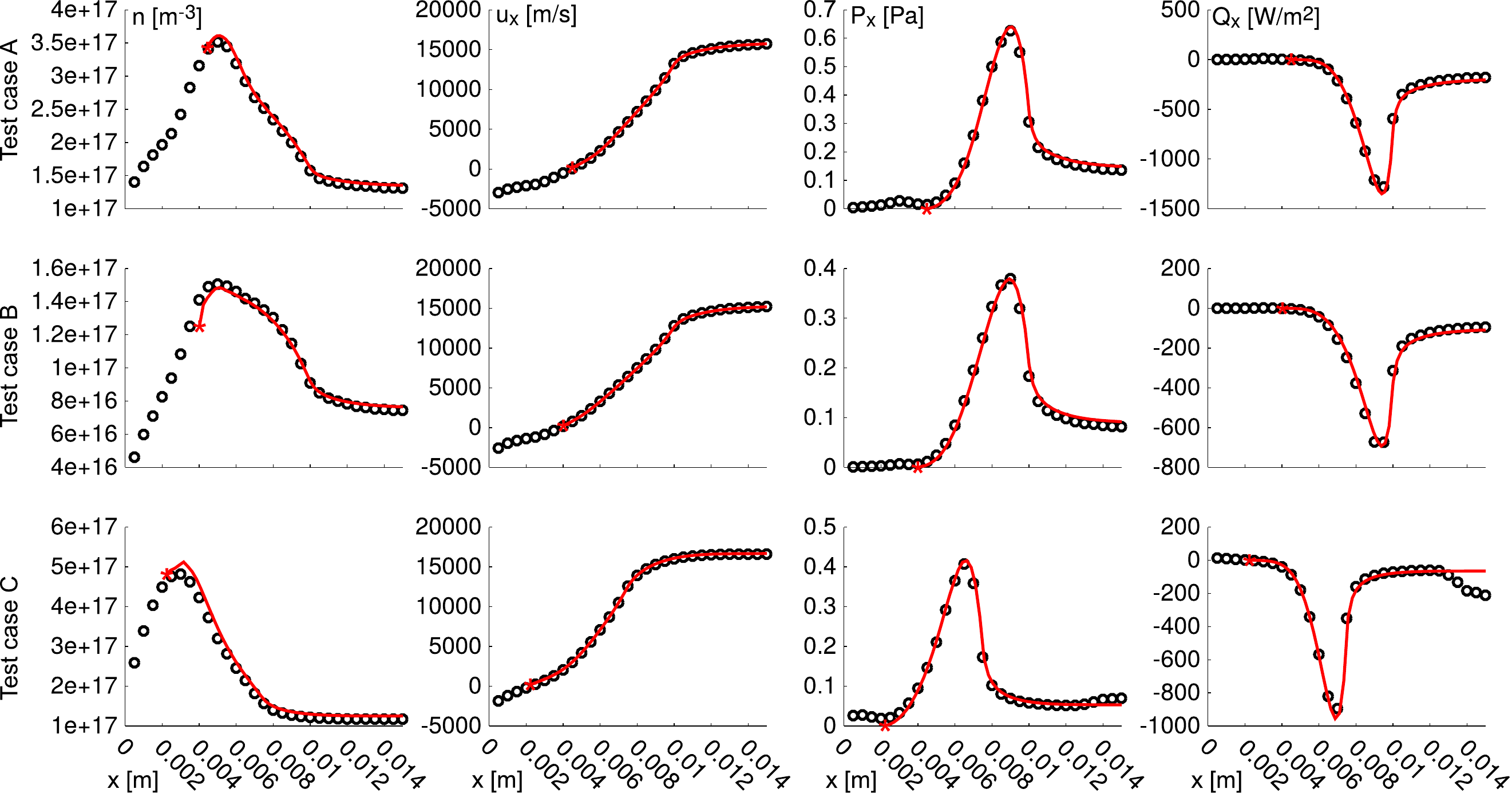}
  \caption{First moments of the ions distribution function for the PIC test cases. Analytical solution ({\protect\redline}); 
           PIC simulation ({\protect\blackcircle \ \protect\blackcircle \ \protect\blackcircle});
           Symbol \textcolor{red}{$*$} denotes the starting point for the integration, taken where $E = 0$.}
  \label{fig:testcases-analytical-PIC-moments}
\end{figure*}

The analytical solution of Eq.~\eqref{eq:analytic-VDF} 
is compared with the PIC test cases in Fig.~\ref{fig:testcases-analytical-PIC-moments}, in terms of first four moments (Eqs.~\ref{eq:analytical-n}--\ref{eq:analytical-Qx}).
The analytical solution is obtained starting the integration from the position $x_0$ taken where the electric field is zero (starting point indicated as ``\textcolor{red}{$*$}'' in Fig.~\ref{fig:testcases-analytical-PIC-moments}).
It is possible to perform the integration in both forward and backward directions, to obtain the solution in the whole domain.
However, this was not done, since we chose to avoid a nonphysical region of PIC simulations, where the ionization profile is artificially imposed to zero, near the anode (see Appendix \ref{sec:appendix-A}).
The ion birth velocity was taken to be $v_n =  0 \ \mathrm{m/s}$.

The analytical and PIC distribution functions are shown in Fig.~\ref{fig:testcase-A-VDF-compar} for Test case A, and provide very similar agreement for Test cases B and C.

The analytical moments show a very good match for all the test cases.
A tiny error can be appreciated near position $x_0$, which is likely due to the hypothesis of mono-energetic (rather than Maxwellian) injection of ions.
This leads to some error in the predicted distribution function, as can be seen in  Fig.~\ref{fig:testcase-A-VDF-compar}.

\begin{figure*}[htpb]
  \centering
  \includegraphics[width=1.0\textwidth]{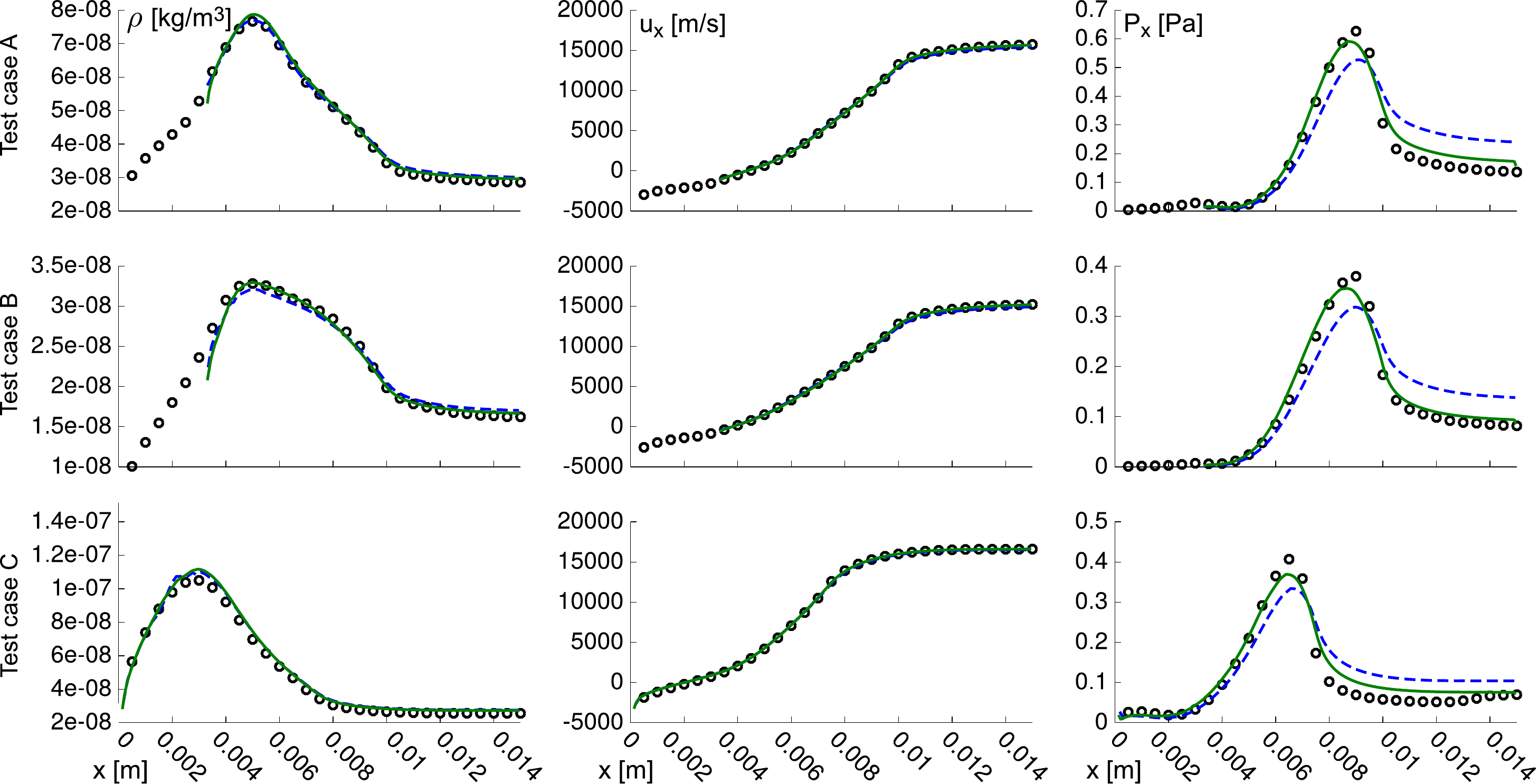}
  \caption{Solution of anisotropic fluid equations for the PIC test cases. 
    PIC simulation ({\protect\blackcircle \ \protect\blackcircle \ \protect\blackcircle});
    anisotropic fluid equations with zero heat flux ({\protect\bluedashedline}); and cubic-VDF heat flux ({\protect\greenlinedark}).}
  \label{fig:testcases-anisotropic-PIC-moments}
\end{figure*}

\subsubsection{Anisotropic equations vs PIC - Test cases A, B, C}\label{sec:results-anisotropic-vs-PIC}


The anisotropic fluid equations
(Eqs.~\ref{eq:conservation-eq-general}, \ref{eq:U-F-anisotr}, \ref{eq:S-anisotr}) were solved for the PIC test cases, imposing the averaged electric field and the ionization profile.
A numerical solution was obtained using a one-dimensional finite volume scheme,\cite{leveque_finite_2002} with second order spatial accuracy (linear one-sided reconstruction of primitive  variables at the interfaces, with the symmetric van Albada slope limiter\cite{van1997comparative} and HLL numerical fluxes).
The problem is solved by marching in time until convergence, with a linearized point-implicit backward Euler scheme. \cite{blazek_computational_2015}
Analogous results were obtained from explicit computations.
A grid composed by 200 cells showed to be fine enough to provide spatial convergence, with the employed second order scheme. Spatial convergence was assessed by performing grid-sensitivity analysis.
The injection velocity and temperature of ions were chosen uniform in space, equal to $v_n = 0 \ \mathrm{m/s}$ and $T_n = 0.5\ \mathrm{eV}$, following the PIC injection conditions.
The computational domain is shown in Fig.~\ref{fig:testcases-E-S}.
For Test cases A and B, the domain was cropped, as to avoid some unphysical behavior of the PIC electric field (see Appendix \ref{sec:appendix-A} for more details).

Results are shown in Fig.~\ref{fig:testcases-anisotropic-PIC-moments}, showing the effect of a simple zero heat flux closure, against the phenomenological $p = 3$ (cubic corrected heat flux) closure developed in this work.
Whereas the zero heat flux closure (anisotropic Euler-like equations) allows to retrieve the density and velocity fields, its accuracy decreases as soon as second order moments such as the pressure are sought.
The $p=3$ heat flux closure on the other hand allows for a strong improvement, despite its apparent simplicity.
The cases of $p = 1$ and $p = 2$ have a somewhat poorer performance.

Some error can be appreciated in the number density for Test case B, around location $x \approx 0.004$ m, which is most likely due to the artificially cropped domain and would disappear in a full simulation.


\subsubsection{Notes regarding collisions and plasma oscillations}

The effect of ion-neutral collisions neglected in this work can be assumed to become of some importance mainly out of the thruster:
electric field and ionization profile quickly go to zero and collisions are the only effect remaining. 

Due to charge exchange (CEX) and momentum exchange (MEX) collisions with neutrals, the ions VDF develops two low velocity structures.\cite{mikellides2005elastic}
Inside the channel and in the near plume, their effect may be neglected at first.
However, progressing along the plume, it quickly becomes important, especially for central moments of second order and higher.
While adding some small low velocity contribution to the VDF does not change the density and changes only slightly the average velocity, the effect becomes much larger for the pressure and heat flux, since the distance between the new contribution and the bulk of the distribution is weighted by a factor $(v - u_x)^p$, with $p$ respectively equal to 2 or 3 for pressure and heat flux.

While we will not consider this in the present contribution, the current description could be extended to include collisions by accounting for multiple families of ions: one population describing the main beam, one for the CEX and one for the MEX ions. 
For each population the set of mass, momentum and energy equations shown here could be solved, coupled by the production term.

The current formulation may also break-down in presence of strong plasma oscillations and ion trapping, whose effect can become important in some circumstances.
This may be the reason of the tiny raise in pressure around location $x = 0.0125$ m for Test case C, becoming a quite visible deviation in terms of heat flux (see Fig.~\ref{fig:testcases-analytical-PIC-moments}).


\subsubsection{Experimental measurements: Test case D}\label{sec:experiment-testcase}

The aim of this test case is to compare the analytical distribution function with all its
simplifications (collisionless, monoenergetic ion birth, steady state) to experimental measurements.
We simulate the conditions of the experiments of Mazouffre et al.\cite{mazouffre_spatio-temporal_2010}
The ionization profile for the considered configuration is taken from Garrigues et al.\cite{garrigues_computed_2012}
This experiment was selected as it provides both the electric field (despite having some acceptable noise) and the 
ionization profile, needed to compute the present solution, together with two distribution
functions for comparison.

\begin{figure}[htpb]
  \centering
  \includegraphics[width=0.95\columnwidth]{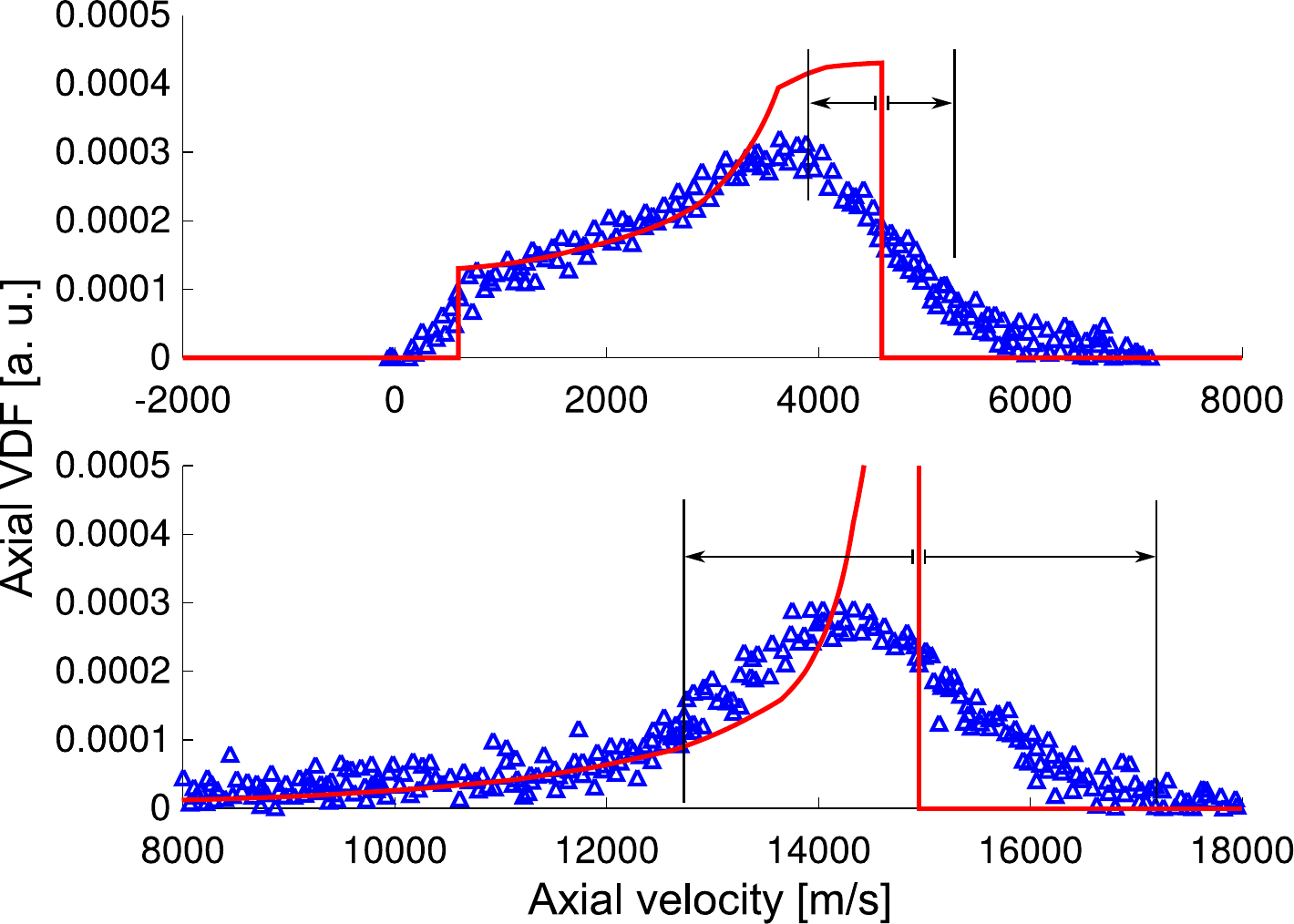}
  \caption{Test case D - Axial VDF. Analytical ({\protect\redline}) vs experimental ({\protect\bluetriangle \ \protect\bluetriangle \ \protect\bluetriangle}). Top: 2 mm inside the thruster, from the exit plane; Bottom: 8 mm after the exit (in the plume).
  Vertical lines: 15\% oscillation.}
  \label{fig:experimental-analytic-VDF}
\end{figure}

We consider the ``current break mode'' of the experimental results.
As no specific detail is given on the velocity of neutral species, we assume an average velocity $v_n = 600\,\mathrm{m/s}$.
This velocity is inferred from the experimental VDF, as it determines the lower velocity tail of the ions VDF 
(Fig.~\ref{fig:experimental-analytic-VDF}-top).
Alternatively, it could be estimated from the mass injection rate, the neutral temperature and the geometrical 
characteristics of the thruster.

Choosing the starting axial position $x_0$ for computing the analytical solution is not straightforward.
The uncertainty on the ion injection velocity adds to the one on the electric field, arising from the measurements noise, as well as from the theoretical method used to reconstruct it.\cite{perez2009method}
This influences phase-space trajectories and the outcome of the analytical solution.
We arbitrarily start from $x = 0.015\ \mathrm{mm}$, which is reasonably close to the position where the electric field becomes positive.

When comparing analytical solution and experimental VDFs, one should consider some additional factors.
First, the temperature of neutrals was supposed to be zero, assuming ions are injected as a monoenergetic beam.
The effect of this assumption can be appreciated in the lower velocity tail of the analytical VDF, which creates 
a sudden jump.
Considering the realistic thermal velocity of 310 m/s for the experiments (suggested by the authors\cite{mazouffre_spatio-temporal_2010}), 
one can explain how the sharp
jump would result in the experimental smoother low velocity tail (where velocity dispersion is indeed in the order of 300 m/s).
Also the sharp high velocity part of the analytical distribution should be expected to smear out in real conditions, partly for the effect of the collisions and partly for the Maxwellian injection process itself.
However, collisions can be shown to have little importance with respect to the accelerating field (and thus generating a deviation which is small with respect to the average velocity), as can be stated from an analysis of characteristic collision times and electrical acceleration,\cite{garrigues_computed_2012} and injection would smear the peak only by some 300 m/s, as mentioned.
Instead, one should consider that the experimental result shows the presence of a breathing mode instability, which following the authors results in a 15\% oscillation of the average velocity (indicatively shown in Fig.~\ref{fig:experimental-analytic-VDF} by vertical bars).
This suggests that in the unsteady case, the high velocity sharp jump of the analytical VDF would oscillate as well, creating a much smoother result.

\begin{figure}[htpb]
  \centering
  \includegraphics[width=0.95\columnwidth]{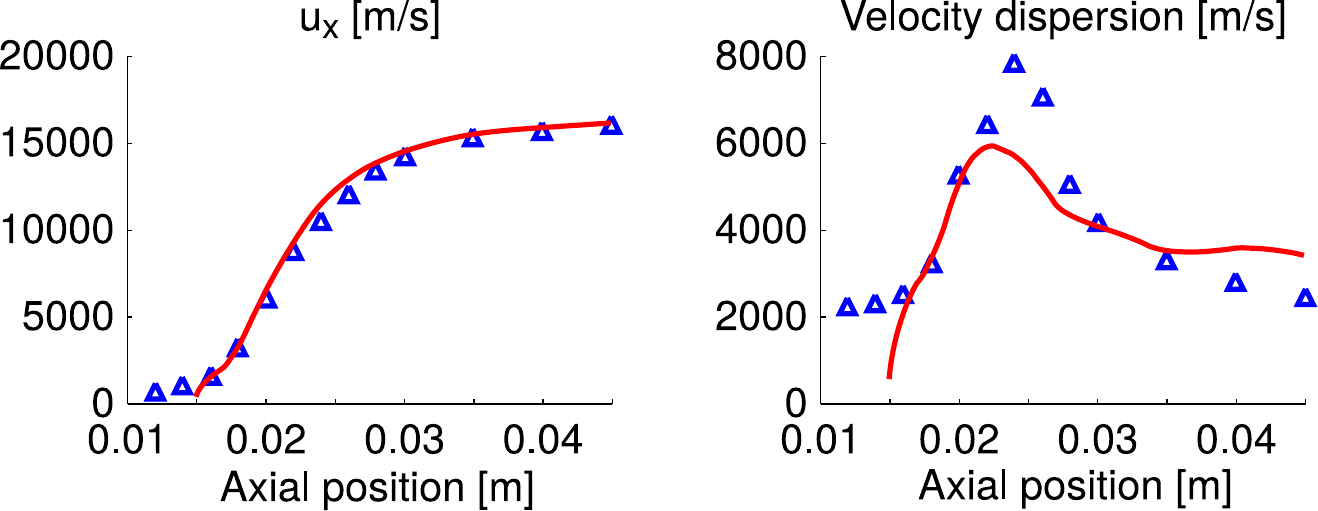}
  \caption{Test case D - Average velocity and velocity dispersion. Analytical (\protect\redline) vs experimental ({\protect\bluetriangle \ \protect\bluetriangle \ \protect\bluetriangle}).}
  \label{fig:experimental-analytic-twomoments}
\end{figure}

The authors provide measurements for the average velocity and the velocity dispersion,\cite{mazouffre_spatio-temporal_2010} defined by Gawron et al.\cite{gawron2008influence}
In Fig.~\ref{fig:experimental-analytic-twomoments}, we compare these values with our analytical results. 
It should be noted that it is quite simple to retrieve a reasonable value for the velocity, while central moments such as the velocity dispersion are much more sensitive, and the electric field oscillations in the considered experiment likely play a large role.
Still, the analytical prediction appears reasonable.

\begin{figure}[htpb]
  \centering
  \includegraphics[width=1.0\columnwidth]{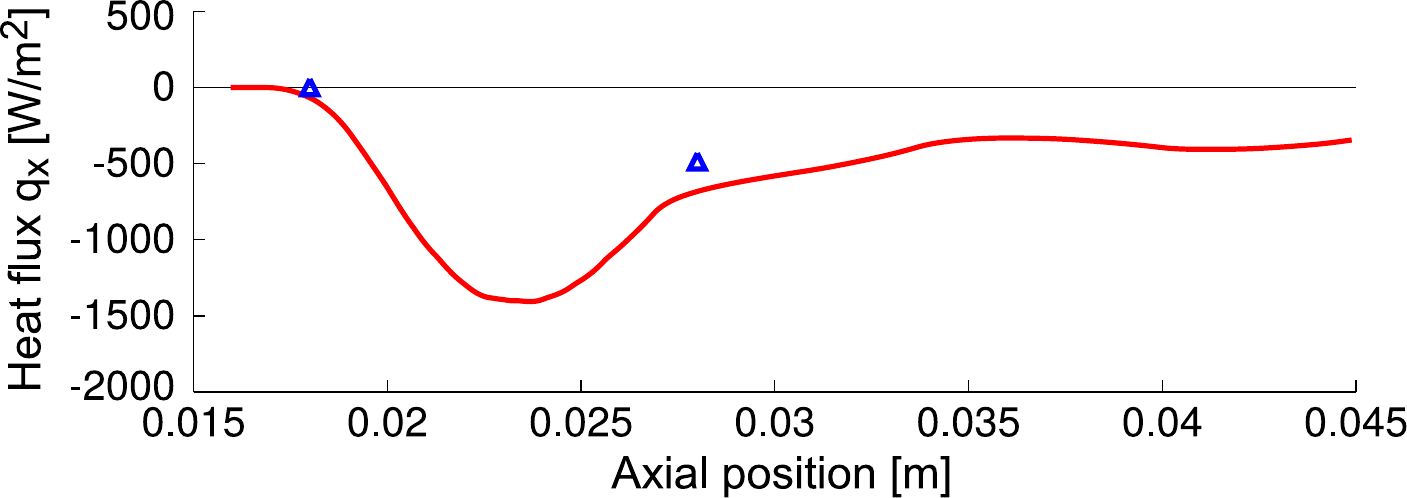}
  \caption{Test case D - Analytical heat flux (\protect\redline) and values reconstructed from experimental VDFs ({\protect\bluetriangle}) of Fig.~\ref{fig:experimental-analytic-VDF}.}
  \label{fig:experimental-analytic-qx}
\end{figure}

Finally, despite the matching of the analytical distribution function and its moments not being perfect, one should consider that the analytical VDF has the strong advantage of being non-Maxwellian, thus providing a non-zero prediction for the heat flux (and the other higher odd moments).
In Fig.~\ref{fig:experimental-analytic-qx} we compare the analytical heat flux with values reconstructed from the two experimental VDFs of Fig.~\ref{fig:experimental-analytic-VDF} at two positions along the channel.
As the experimental VDFs are obtained from LIF and provided in arbitrary units, the comparison is done scaling them as to give the same number density as the analytical result.
Unfortunately, the lack of data does not allow for a real validation of the heat flux.
However, the two available points indicate a partial agreement.
Moreover, since the heat flux is a central moment, we could expect a degree of accuracy smaller than that of the average velocity (Fig.~\ref{fig:experimental-analytic-twomoments}-Left), but qualitatively analogous to the one obtained for the velocity dispersion (Fig.~\ref{fig:experimental-analytic-twomoments}-Right).

\section{Conclusions} \label{sec:conclusion}

In this work we focused on the description of the axial behavior of positive ions in Hall thruster discharges, assumed to be fully collisionless.
Neutrals and electrons have not been included in the description, as well as the ions azimuthal and radial behavior.
This limits the validity of the discussed model to the region near the channel centerline, and makes the model predictive in an azimuthally-averaged sense.

Since ions can be considered to be low-collisional, the results from this work can be readily supplemented by equations describing the remaining species and the two directions excluded from the analysis.

First, the axial behavior of ions has been discussed from the standpoint of kinetic theory.
The formation of the axial ion velocity distribution function (VDF) was described by analyzing the ionization profile and axial electric field.
Ions that are created inside the channel, where the electric field is low, accumulate in phase-space and constitute the characteristic peak in the ions axial velocity distribution function.
Ions that are created later on, where the electric field is larger in magnitude, compose the body, or heavy tail of the ions VDF.
Most of the velocity dispersion is created in the region where electric field and ionization profile overlap, as remarked by Mazouffre and Bourgeois.\cite{mazouffre_spatio-temporal_2010}

A simple analytical VDF was obtained by establishing a balance of fluxes in phase-space, following the formulation for plasma sheaths.
The analytical result assumes steady state and monoenergetic creation of ions. Despite these assumptions, the analytical result proved able to match closely collisionless Particle-In-Cell simulations in terms of distribution function and its moments, even in presence of oscillations most likely related to electron drift instabilities.
The analytical model proposed in this work could be employed to post-process experimental results, or as an accurate modeling tool for describing ions in one-dimensional and steady state simulations.
The solution could be also applied in quasi-steady state conditions, in the case where the electric field and ionization profiles vary slowly with respect to the residence time of ions inside the channel. 
This may be the case for certain low-frequency oscillations.

Moments from the analytical solution were obtained as integral expressions of the electric field and ionization profile over the domain.
A comparison with collisionless Particle-In-Cell simulations showed an almost exact matching.

With the aim of providing an accurate and cost-effective description that is able to reproduce the axial kinetic behavior of ions, an anisotropic fluid formulation was derived by integration of the one-velocity Vlasov equation.
Such model needs a closure and even though a simple adiabatic closure ($Q_x = 0$, corresponding to anisotropic Euler equations) proved able to 
retrieve the correct lower order moments (density and velocity profiles), it exhibited a significant error in the prediction of the pressure and temperature, which are second order central moments.

Therefore, we developed a phenomenological closure based on the approximation of the distribution function by either a triangular, parabolic or cubic function.
This assumption leads to an analytic closure, giving the heat flux in function of its lower order moments: density and temperature.
The heat flux obtained with a cubic approximation of the VDF showed able to bring a significant improve over the simple adiabatic closure, reproducing well the Particle-In-Cell heat flux and pressure profiles.

The system of ions anisotropic equations developed in this work can be readily inserted into a larger framework, where equations for neutrals and electrons are solved at the same time, and are coupled to the ions equations through the electric field and ionization profile.\cite{barral_numerical_2001,martorelli2019comparison}
Moreover, the present fluid framework could be extended to multiple space dimensions.
This would consist in adding additional momentum and temperature equations for each considered dimension. A closure for the newly introduced pressure tensor and heat flux components would also need to be carefully chosen.
As a very first approximation, one could for example assume a Maxwellian distribution function in the azimuthal and radial directions, therefore obtaining a simplified pressure tensor and zero heat flux in such directions, while still including the currently proposed form for its axial component.

The proposed anisotropic fluid model goes in the direction of providing fluid descriptions with enhanced accuracy over classical Euler or Navier-Stokes-Fourier formulations.
Once properly extended to higher dimensions, the proposed model could be employed in accurate fully-fluid simulations, which could maybe reach analogous accuracies as the more complicated kinetic-fluid hybrid models.

Finally, we should remark that the assumption of fully collisionless ions is reasonably valid inside the thruster and in the near plume, where electrostatic acceleration dominates the ions dynamics, but progressively loses accuracy along the plume, where in absence of strong electric fields, collisions are the only effect left.

The data that supports the findings of this study are available within the article.


\section*{Acknowledgments}
The authors wish to acknowledge prof. A.~Frezzotti (Politecnico di Milano), prof. J.~McDonald (University of Ottawa) and prof. C.~Groth (UTIAS Toronto), for the useful discussions, as well as Mr. G.~Gangemi.


\appendix
\section{PIC Test cases A, B, C}\label{sec:appendix-A}

As a mean to assess the quality of our results, we have performed three Particle-In-Cell Monte-Carlo-Collisions (PIC-MCC) simulations, with the code \textit{LPPic} that was already verified via the 1D Helium benchmark of Turner et al. \cite{Turner2013} and extensively used to simulate the radial-azimuthal plane of a Hall Thruster.~\cite{Croes2017,Tavant2017} The code was adapted to simulate the axial-azimuthal plane and we present here three cases with a simulation model similar to the one used in the 2D benchmark of Charoy et al.\cite{Charoy2019}

\begin{figure}[htpb]
  \centering
  \includegraphics[width=0.9\columnwidth]{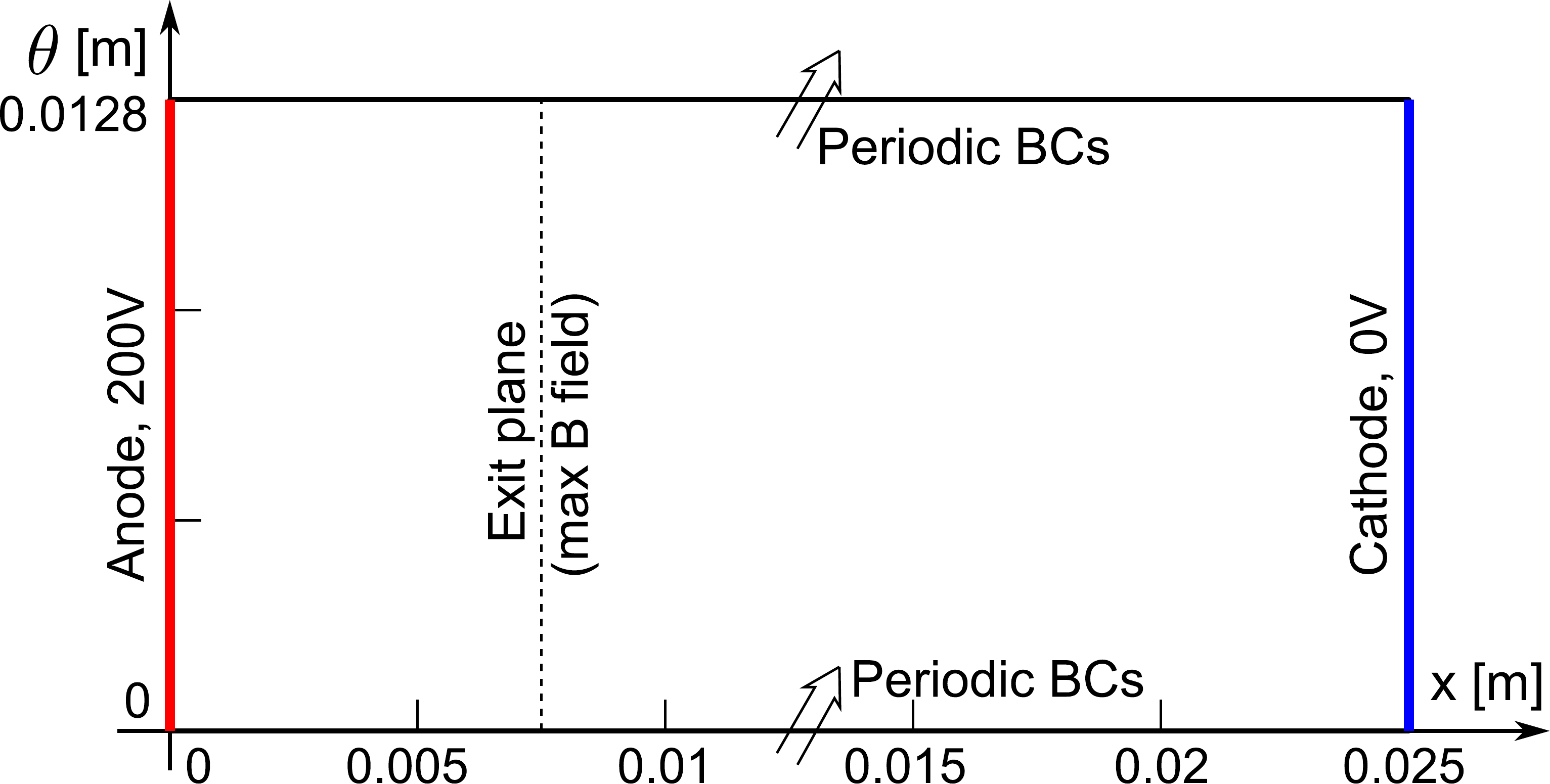}
  \caption{Computational domain for PIC simulations of Test cases A, B and C. Dashed line marks the position of maximum magnetic field $B$.}
  \label{fig:PIC-schematic}
\end{figure}

In this case, an axial electric field is created by a potential difference imposed between an anode at 200 V and a cathode at 0 V (see Fig.~\ref{fig:PIC-schematic}). The azimuthal direction is considered as periodic with a fixed length of 1.28 cm while 2.5 cm are simulated in the axial direction. To comply with PIC stability constraints, we used a time step of $\Delta t = 5 \times 10^{-12}$ s and a cell size of $\Delta x = 5 \times 10^{-5}$ m. As this case is collisionless, an ionization profile is imposed, leading to the injection of ion/electron pairs along the 
channel with a given initialisation temperature of $T_i$=0.5 eV and $T_e$ = 10 eV. Electrons are injected at the cathode line to sustain the discharge. 
The imposed magnetic field reaches its maximum at position $x = 0.0075$ m, which we assume to represent the exit plane of the thruster.
The reader can refer to the aforementioned paper for more details. 

Test case A is exactly identical to the case of Charoy et al.\cite{Charoy2019} with an imposed ion current density of $400 \ \mathrm{A/m^2}$.
Test case B is done with a lower ion current density of $200 \ \mathrm{A/m^2}$, which leads to a lower ionization profile.
Test case C is similar to test case A, but the ionization profile has been shifted towards the anode, in order to artificially generate a longer acceleration region.

\begin{figure}[htpb]
  \centering
  \includegraphics[width=1.0\columnwidth]{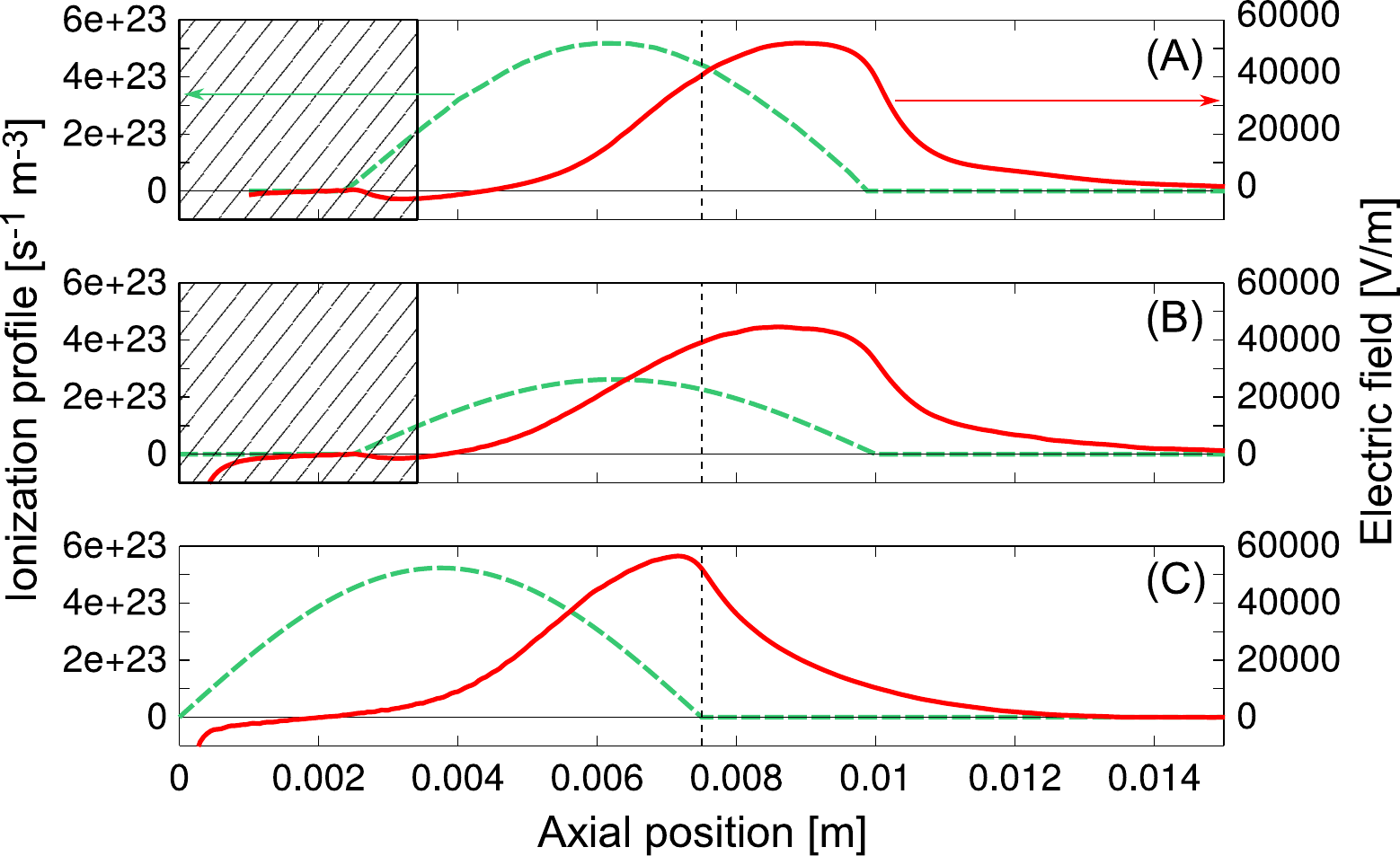}
  \caption{Profiles of ionization and electric fields for the PIC Test cases A, B and C. Hatched regions show domain excluded from the fluid computations. Dashed line is the location of maximum magnetic field in PIC simulations (thruster exit plane).}
  \label{fig:testcases-E-S}
\end{figure}

The simulations performed show azimuthally travelling waves, characteristic of the electron drift instability.
The PIC fields shown in this work were obtained by averaging along the azimuthal direction, at different axial locations. To reduce the statistical noise and to filter oscillations arising from azimuthal instabilities, a time average is performed on 40 samples, spaced in time by 5000 time steps as to provide statistical independence and to sample adequately during the oscillation period. An analysis of different time-averaging settings showed no sensible variation of the average fields, except from statistical noise.

The resulting axial electric field profile is shown in Fig.~\ref{fig:testcases-E-S} for the 3 test cases, together with the imposed ionization profiles.
These fields are the required inputs for the analytical solution and the 1D anisotropic fluid equations.

While the hypothesis of imposing an ionization profile is a good way of getting reasonable steady results, the cosine shape employed may generate some oscillations or nonphysical behaviors in the PIC simulations, especially near the anode, where the amount of ions is low.
In some simulations, this can be appreciated as tiny oscillations of the averaged electric field, (Test cases A and B at position $x \approx 0.0025$ m), which quite likely do not have a physical origin.
Therefore, for the analytical and anisotropic fluid simulations of Test cases A and B, we focus on a restricted region of the original PIC simulations, shown in Fig.~\ref{fig:testcases-E-S}.

This does not constitute a limitation for the current work, where we apply these fields directly into the anisotropic equations: this unphysical behavior would be relaxed once the equations developed are inserted into a fully coupled multi-fluid formulation, where the ionization profile is not artificially imposed and the Poisson equation solved self-consistently.

Also, in the fluid simulations, we simulate only up to the axial position $x = 0.015$ m. 
Fluid simulations are not conducted further in order to stay far from the cathode region of the PIC simulation, positioned at $x = 0.025$ m.
Indeed, for these 2D simulations, the cathode is modeled in such a way that an artificial sheath is created , which does not represent properly the potential distribution generated by a cathode in a real 3D scenario.
Limiting the fluid simulation domain to $x = 0.015$ m is enough to capture the electrostatic acceleration and the ions production, in which we are interested in the present work.


\section{Maxwellian ions injection}\label{sec:appendix-B}

\begin{figure}[htpb]
  \centering
  \includegraphics[width=1.0\columnwidth]{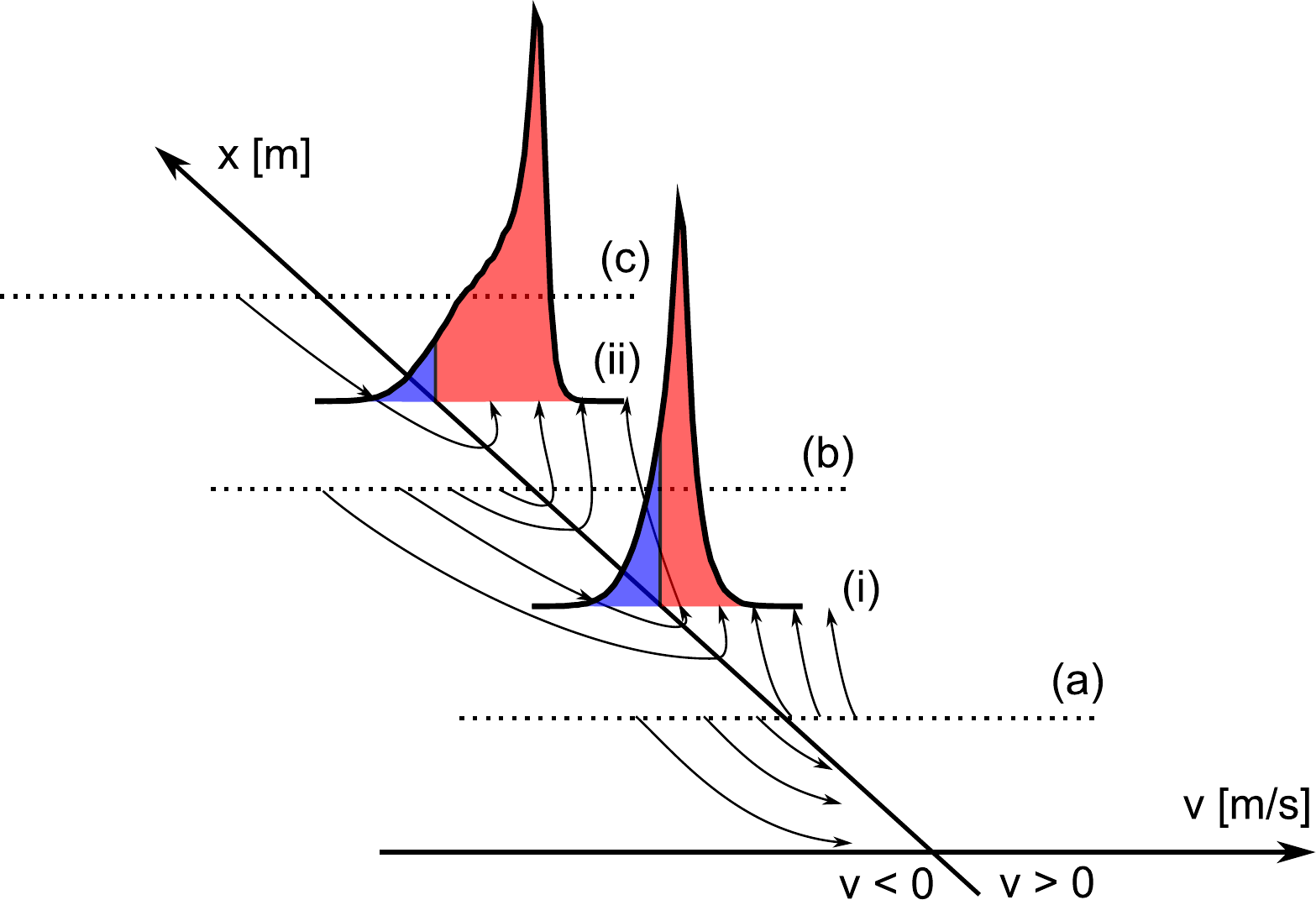}
  \caption{Ions path in phase-space in a positive electric field region.}
  \label{fig:maxwellian-injection}
\end{figure}

An accurate prediction of the ions VDF in the low velocity region requires accounting for the birth distribution function of ions.
Considering the axial position (i) in Fig.~\ref{fig:maxwellian-injection}, denoted by $x_i$, and considering a positive electric field for simplicity, the VDF appears constituted by three contributions.
(1) ions created with positive velocity at positions before $x_i$, contribute to the (red) right side of the ions VDF;
(2) ions created with negative velocity and after the position $x_i$ backstream towards $x_i$ and contribute to the (blue) negative velocity side of the VDF;
Finally (3), these same ions are slowed down by the electric field until they gain positive velocity, and ultimately contribute to the positive side of the ions VDF.
These contributions are to be included if the detailed effects of a Maxwellian injection are sought.
However, assuming a monoenergetic injection is probably enough from the point of view of the acceleration region.

\section{Closure for order $p$ polynomials}\label{sec:appendix-C}

This appendix reports the heat flux closure for polynomial distribution functions of order $p$ (not necessarily integer). 
From the density $n$, one obtains the relation:
\begin{equation}
a = n\ \frac{p+1}{L^{p+1}}
\end{equation}

\noindent The velocity extremes are obtained from the momentum as:
\begin{equation}
\begin{cases}
V_A = u - \frac{p+1}{p+2}L \\
V_B = u + \frac{1}{p+2}L
\end{cases}
\end{equation}

\noindent The ``width'' $L$ of the polynomial distribution is obtained as:
\begin{equation}
L = \sqrt{\frac{k_B T}{m} \left[ \frac{p+1}{p+3} - \left(\frac{p+1}{p+2}\right)^2 \right]^{-1} }
\end{equation}

\noindent And the heat flux closure results in:
\begin{equation}
Q_x = \frac{m n}{2} L^3 \left[ \frac{p+1}{p+4} - 3 \frac{(p+1)^2}{(p+2)(p+3)} + 2 \left(\frac{p+1}{p+2}\right)^3\right]
\end{equation}

A regularization for negative velocities and a limiting around zero velocity 
should then be applied as detailed in the text.

%
\nocite{*}
\bibliography{ions_VDF}{} 
\bibliographystyle{plain}

\end{document}